\documentclass[12pt,preprint]{aastex}

\usepackage{epsfig}
\usepackage{amsmath}
\usepackage{amssymb,amsfonts,textcomp}
\usepackage{color}
\usepackage{subfigure}
\usepackage{longtable}

\shorttitle{Dedispersion Power Analysis}
\shortauthors{Clarke, Macquart \& Trott}

\begin{document}


\title{Performance of a novel fast transients detection system}


\author{Nathan Clarke} \email{N.Clarke@curtin.edu.au}
\author{Jean-Pierre Macquart\altaffilmark{1}}
\and
\author{Cathryn Trott\altaffilmark{1}}
\affil{ICRAR/Curtin University, Bentley, WA 6845, Australia}


%


\altaffiltext{1}{ARC Centre of Excellence for All-Sky Astrophysics (CAASTRO)}

\begin{abstract}
We investigate the S/N of a new incoherent dedispersion algorithm optimized for FPGA-based architectures intended for deployment on ASKAP and other SKA precursors for fast transients surveys.  
Unlike conventional CPU- and GPU-optimized incoherent dedispersion algorithms, this algorithm has the freedom to maximize the S/N by way of programmable dispersion profiles that enable the inclusion of different numbers of time samples per spectral channel.  This allows, for example, more samples to be summed at lower frequencies where intra-channel dispersion smearing is larger, or it could even be used to optimize the dedispersion sum for steep spectrum sources.
Our analysis takes into account the intrinsic pulse width, scatter broadening, spectral index and dispersion measure of the signal, and the system's frequency range, spectral and temporal resolution, and number of trial dedispersions.
We show that the system achieves better than 80\% of the optimal S/N where the temporal resolution and the intra-channel smearing time are smaller than a quarter of the average width of the pulse across the system's frequency band (after including scatter smearing).
Coarse temporal resolutions suffer a $\Delta t^{-1/2}$ decay in S/N, and
coarse spectral resolutions cause a $\Delta\nu^{-1/2}$ decay in S/N,
where $\Delta t$ and $\Delta\nu$ are the temporal and spectral resolutions of the system, respectively.
We show how the system's S/N compares with that of matched filter and boxcar filter detectors.  
We further present a new algorithm for selecting trial dispersion measures for a survey that maintains a given minimum S/N performance across a range of dispersion measures.  

\end{abstract}

\keywords{methods: observational --- surveys --- instrumentation: detectors --- pulsars: general --- radio continuum: general}

\section{Introduction}

The dispersive nature of the plasma that pervades interstellar and intergalactic space causes the observed arrival time of impulsive astrophysical radio signals to be strongly frequency dependent.  In cold plasmas the dispersive delay is proportional to $\lambda^2 \, \text{DM}$, where the dispersion measure, DM, is the line-of-sight electron column density.  The effects of dispersion are particularly manifest in searches for pulsars and short-timescale transients at long wavelengths ($\lambda \gtrsim 0.1\,$m) with sufficient sensitivities to detect objects at large distances.  This applies to several current and planned high-sensitivity surveys on next-generation radio telescopes, which are being conducted in the regime in which the effects of interstellar, and potentially intergalactic, dispersion are extreme (e.g.~the LOFAR Transients Key Project; \citealt{2011A&A...530A..80S}; CRAFT, \citealt{macquart2011}; Arecibo PALFA Survey; \citealt{cordes2006}; HTRU survey \citealt{keith2010}; \citealt{burke_spolaor2011}).

 

The effects of dispersion smearing are in principle fully reversible if the electron column through which the radiation propagated can be determined.  However, a number of practical factors prevent complete recovery of the signal to the same strength as an undispersed pulse.  For the process of incoherent dedispersion, in which the signal is reconstructed from a filterbank of intensities gridded in time and frequency \citep{2003ApJ...596.1142C}, there are three primary means by which the S/N is degraded.  1. The finite resolution of the filterbank limits the S/N of the dedispersed signal when there is residual dispersion smearing across the individual filterbank channels (i.e.~when the dispersive delay across the bandwidth of the channel exceeds the temporal resolution).  2. Finite computational power limits the number of DM trials that can be searched in a survey, resulting in a loss of sensitivity to events with DMs in between trials.  3.  The signal is smeared over a large number of temporal bins, which degrades the signal strength in the presence of system noise \citep{2003ApJ...596.1142C}.

The process of coherent dedispersion \citep{hankins-sp1975}, in which the raw signal voltages recorded from the antenna are convolved with the inverse of the transfer function of the dispersive medium, achieves the optimum S/N recovery of the dispersed signal by eliminating effects 1 and 3.  However, for the purposes of conducting blind surveys for one-off transient events, the data- and compute-intensive nature of coherent dedispersion renders it too slow to be practical with present technology.

The technique of incoherent dedispersion offers a viable alternative when processing resources are limited. Incoherent dedispersion is the mainstay of most current pulsar search and transients survey detection algorithms \citep[e.g.,][]{2011ApJ...735...97W,terveen2011}.  A complete understanding of its performance is crucial to understanding the optimal dedispersion strategy when computational resources are finite.  For instance, if a real-time detection system can only dedisperse the signal at a fixed number of trial dispersion measures, what is the optimal choice of trial DMs?  A related problem is to quantify the effect of a given dedispersion strategy on the completeness statistics of the survey.  Though these are old questions, the answers have acquired a renewed urgency because they are needed to inform the design of next generation surveys for impulsive signals \citep[e.g.,][]{daddario-searching2010}.  These questions have been addressed in the past \citep[e.g.,][]{2003ApJ...596.1142C}, but without addressing the degrading effects of implementing boxcar templates as opposed to true matched filters, and only considering a general approach to analysing the effects of temporal and spectral resolution, and DM error.  The optimization of blind surveys for pulsars and transients is particularly pressing in the context of SKA time-domain system design, where extreme data rates make offline data storage impractical in many instances, and necessitate real-time processing of the data stream.   These factors influence SKA system design and drive backend hardware processing requirements, which can comprise a sizable fraction of the total cost of the instrument.

Incoherent dedispersion techniques have been employed for several decades.  An early technique, known as the tree algorithm \citep{taylor1974sensitive}, consists of a regular structure of delay and sum elements that transforms an input signal of $N$ frequency channels to $N$ dedispersed output signals, with $O(N\log_2 N)$ operations.  While the tree algorithm is a process-efficient technique and has been popular, particularly in early pulsar surveys, it has some draw-backs that limit its sensitivity: a) it assumes that signal dispersion is linear with frequency, b) the dispersion measures for each of the dedispersed outputs are fixed to linear distributions from 0 (no dispersion) to the DM at which the gradient of the dispersion curve is one temporal bin per spectral channel (thus called the ``Diagonal DM''), and c) each dedispersed output sample is the sum of only one sample from each of the $N$ channels of the dynamic spectrum.  Additional processing stages are often employed to mitigate some of these limitations: for example, \citet{2001MNRAS.328...17M} linearize dispersion by inserting artificial (``dummy'') channels between the real frequency channels, and then divide the linearized data into smaller groups of adjacent channels, or sub-bands, before dedispersing each sub-band using the tree algorithm; and a broad distribution of trial DMs is achieved by successively summing the data samples in pairs and repeating the dedispersion process.

Another algorithm called DART (a Dedisperser of Autocorrelations for Radio Transients) used in the V-FASTR transient detection system for the VLBA \citep{2011ApJ...735...97W} arranges samples of the signal's dynamic spectrum into vectors, one vector per frequency channel, with each vector containing a time series of samples of up to several seconds.  The vectors are then skewed with delay offsets appropriate to the trial DM, then summed to produce the dedispersed time series for that trial.  In many ways the DART algorithm is more flexible than the tree algorithm: It supports an arbitrary number and distribution of trial dispersion measures, and it supports arbitrary dynamic-spectrum dispersion curves, including curves proportional to $\lambda^2$.  However, it too sums only one sample from each input channel to produce each dedispersed output sample.


A new transients detection system called Tardis is being developed for the Commensal Real-time ASKAP Fast Transients (CRAFT) survey \citep{2010PASA...27..272M}.  For this system \citet{daddario-searching2010} describes a dedisperser that can, for each output sample of a given trial, sum dynamic spectrum samples from multiple temporal bins per spectral channel.  Thus, for large DMs where pulse power can be distributed over many temporal bins per spectral channel, additional dynamic spectrum samples can be included in the sum to improve the S/N of the dedispersed output.  The Tardis implementation of this system (Clarke et al., in prep.) allows arbitrary sets of dynamic spectrum samples to be selected for the dedispersion sums for each trial.  The samples of each set are selected a priori depending on the DM, pulse width and spectral index assumed for the trial.  The pulse width can include the signal's intrinsic width and also temporal broadening of the signal due to interstellar scattering.  Equal weight is given to all samples in each trial sum.

In this paper, we examine the S/N performance of the fast transients detector proposed in \citet{daddario-searching2010} and implemented in Tardis, and we present a sample selection algorithm aimed at maximizing the S/N of each dedispersed output signal.  While matched filter detectors perform weighted sums of signal samples, with weightings determined by assumed pulse profiles, we show that our new detector yields comparable performance using unweighted sums.   In the second part of the paper, we use the new detector to describe how performance is affected by the temporal and spectral resolutions of the system, the magnitude of dispersion and DM error. We use these results as tools with which to decide how to choose the optimal balance of resources for a given system (spectral and temporal resolution, and trial DMs), extending previous work in these areas to form concrete recommendations for system design with dynamic spectrum detectors.


In \S\ref{sec:overview} we define the problem and specify the Tardis dedispersion algorithm mathematically.
The S/N reduction associated with finite temporal and spectral resolution is examined in \S\ref{sec:timefreq}, and in \S\ref{sec:snr_vs_dm_approx} we examine how the S/N reduces with increasing dispersion measures.  
In \S\ref{sec:comparison} we compare the performance of the new algorithm with that of time-series and dynamic spectrum matched filters, and the traditional boxcar filter.
Then in \S\ref{sec:completeness} we study the residual temporal smearing due to differences between trial DMs and true dispersion measures of signals (i.e.~DM errors), how these errors impact the S/N performance, and present a new algorithm for selecting trial DMs to maximize the completeness of fast transients surveys.  Our conclusions are outlined in \S\ref{sec:conc}.




\section{A dynamic spectrum fast transients detection system} \label{sec:overview}

In this section we examine the S/N performance of the incoherent fast transients detection system outlined in \citet{daddario-searching2010} and advance an alternative sample selection algorithm that aims to maximize the S/N performance.

\subsection{Dedispersion fundamentals} \label{sec:defns}

Consider a pulse whose intrinsic emitted power per unit bandwidth is of the form,
\begin{eqnarray}\label{eqn:intrinsic_power}
P_\nu (t,\nu) = P_0 \left( \frac{\nu}{\nu_0} \right)^{-\alpha} f(t),
\end{eqnarray}
where $P_0$ has dimensions W\,Hz$^{-1}$, $\alpha$ is the spectral index of the pulse and $f(t)$ is a dimensionless function that describes the intrinsic pulse profile.  $P_\nu$ is the energy received per unit time per unit bandwidth at a given time $t$ and frequency $\nu$.\footnote{Formally $P_\nu$ cannot be treated as a continuous function of both time and frequency to arbitrary precision in both quantities, since time and frequency are dual parameters connected via the Fourier transform.  However in practice, in the regime where $\Delta\nu\,\Delta t \gg 1$, eq.~(\ref{eqn:intrinsic_power}) is an excellent approximation to a continuous function because the discretization is on a much finer scale. 
}

Interstellar dispersion introduces a delay in the signal arrival time of an amount $t_d = \text{DM}/{\kappa \nu^2}$, where the dispersion measure (DM) is the integral of the electron density along the propagation path of the signal, and $\kappa = 2.41\times10^{-16}$ pc.cm$^{-3}$.s is a constant \citep{hankins-sp1975}.
Furthermore, multipath propagation, or scattering, in the interstellar medium can cause broadening of the temporal width of the signal, and diffractive and refractive scintillation modulations of the signal intensity \citep{1990ARA&A..28..561R}.  

Scattering is highly dependent on the signal frequency, and on the direction and distance of the source in a manner that strongly correlates with dispersion measure.
We model scatter broadening as a convolution in time (denoted by an asterisk) with a general scattering impulse response function, $h_d\left(t;\nu,\text{DM}\right)$.  $h_d$ is dimensionless and as temporal smearing due to scattering involves no attenuation in signal power, its area is unity.  ($h_d$ approaches the dirac delta function in the limit of no scattering.)

Scintillation causes deep (up to 100\% of the mean) amplitude modulations in time and frequency.  Scintillation time scales are generally too large to be relevant to detecting fast transients.
The only instance in which frequency modulation plays an important role is where the decorrelation bandwidth is comparable to the observed bandwidth; larger modulations affect all frequencies within the observed bandwidth equally, and smaller modulations average-out across the band.
Optimization of the S/N subject to the effects of scintillation is prohibitive in a computationally limited system, because scintillation is a stochastic process with multitudes of possibilities that compound an already large parameter space.  For this reason we choose not to include scintillation in our model.

Considering dispersion and temporal smearing due to scattering, our model for the observed power per unit bandwidth is
\begin{eqnarray}\label{eqn:P_obs}
P_{\nu,\text{obs}} (t,\nu) = P_0 \left( \frac{\nu}{\nu_0} \right)^{-\alpha} f \left( t - \frac{\text{DM}}{\kappa \nu^2} \right) * h_d\left(t;\nu,\text{DM}\right).
\end{eqnarray}
The average power received over temporal and spectral intervals $[t,t+\Delta t]$ and $[\nu,\nu+\Delta \nu]$ respectively is
\begin{eqnarray}\label{Pavg}
\bar{P}(t,\nu) =  \frac{1}{\Delta t} \displaystyle\int_{\nu}^{\nu + \Delta \nu} d\nu' \displaystyle\int_t^{t+\Delta t} dt' \, P_{\nu,\text{obs}} (t',\nu').
\end{eqnarray}

\begin{figure}[ht]
  \framebox[\textwidth]{\includegraphics[scale=1.5]{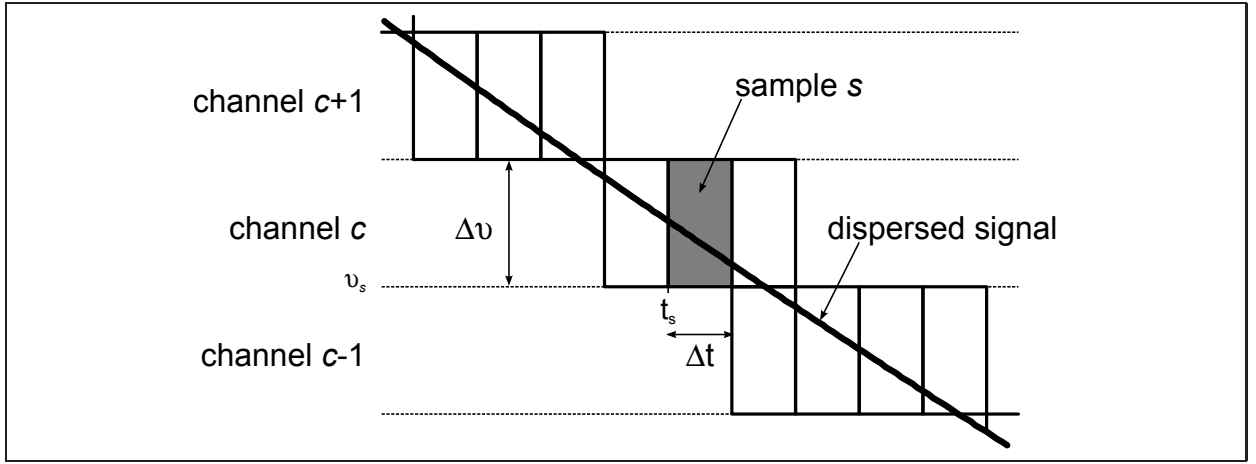}}
  \caption{\label{fig:sample_pwr}Defining points of interest in determining the dedispersed output time series for a trial.}
\end{figure}

In digital systems, the dynamic spectrum of a signal is quantised in frequency and time into discrete samples.  If we assume that the time dimension is quantised to a resolution of $\Delta t$ and that frequency is quantised into channels of $\Delta \nu$, then each sample represents the average power within a $\Delta t$-by-$\Delta \nu$ cell of the dynamic spectrum, as illustrated in Figure~\ref{fig:sample_pwr}.  Each sample includes contributions from the signal, i.e.~the dispersed pulse, and noise from the sky and the receiver.  Thus if sample $s$ represents the average power in the cell $[t_s,t_s+\Delta t;\nu_s,\nu_s+\Delta \nu]$, 
then sample $s$ would have a value of ${\bar P}(t_s,\nu_s) + {\bar P_N}(t_s,\nu_s)$, where the former term is the average power of the pulse within the cell (as modeled in eq.~(\ref{Pavg})), and the latter term is the average noise power within the cell.

The system described by \citet{daddario-searching2010} involves summing selected samples of the dynamic spectrum, where samples are selected based on their relative time $t$ and frequency $\nu$, and on the dispersion measure, pulse width and spectral index assumed for the trial.  We will consider how to select the samples in the next section.  For now, assume that $\mathbb{S}$ is the set of samples selected to dedisperse the signal for a given trial.  The pulse component (ignoring noise) of the time series output of the dedisperser for that trial can be modeled as
\begin{equation}\label{eqn:p_dedisp}
  P_{dedisp}[n] = \displaystyle\sum\limits_{s\in\mathbb{S}}{\bar P}(t_s+n\,\Delta t,\nu_s),\quad\forall n\in\mathbb{Z}.
\end{equation}




For the purposes of detecting astronomical pulses, we aim to maximize the dedispersed signal power relative to statistical variations in the noise power.  Our figure of merit is therefore the signal-to-noise ratio (S/N) calculated as a ratio of $P_{dedisp}$ to the noise error (i.e.~the standard deviation of the noise).
The uncertainty principle implies that the product of the temporal and spectral resolution cannot be less than unity, and in this paper we assume that $\Delta \nu\,\Delta t \gg 1$ such that the central limit theorem holds and the noise contribution to each sample can be assumed to be normally distributed.  To simplify our analysis, we ignore self-noise generated from the signal; self-noise is typically small compared with sky and receiver noise.  Using the radiometer equation, the noise error in a cell of bandwidth $\Delta\nu$ and interval $\Delta t$ can be modeled as
\begin{equation}
  \sigma_n = \frac{k\,T_{sys}\,\Delta\nu}{\sqrt{\Delta\nu\,\Delta t}},
\end{equation}
where $k$ is Boltzmann's constant and $T_{sys}$ is the system equivalent noise temperature.
Generally, $T_{sys}$ is a frequency dependent parameter that represents the overall noise temperature of the system, including natural radio emissions from the sky and gain fluctuations in the receiver electronics; however, variations in system temperature are often relatively small across the operating bandwidth of the receiver, and for the purposes of the analyses in this paper we assume that $T_{sys}$ is constant with frequency.
Since the noise is normally distributed, the average total noise power after summing the samples for a given trial is
\begin{equation}\label{eqn:P_N_dedisp}
  \sigma_{n_{dedisp}} = \sqrt{\displaystyle\sum\limits_{s\in\mathbb{S}}\sigma_n^2} = \sqrt{N_\mathbb{S}}\,\frac{k\,T_{sys}\,\Delta\nu}{\sqrt{\Delta\nu\,\Delta t}},
\end{equation}
where $N_\mathbb{S}$ is the number of samples in set $\mathbb{S}$.  The dedispersion process therefore produces a S/N ratio given by:
\begin{eqnarray}\label{eqn:snr}
  \text{SNR}[n] = \frac{P_{dedisp}[n]}{\sigma_{n_{dedisp}}}
  = \sqrt{\frac{\Delta t}{N_\mathbb{S}\,\Delta\nu}}\,\frac{\displaystyle\sum\limits_{s\in\mathbb{S}}{\bar P}(t_s+n\,\Delta t,\nu_s)}{k\,T_{sys}}.
\end{eqnarray}

\subsection{Sample selection for maximum Signal-to-Noise Ratio (S/N)}\label{sec:selection}

Assume that we have a set of samples $\mathbb{S}$ for dedispersing our signal, and consider the possibility of adding another sample, $\varsigma$, to our set.  If we were to include this sample, then the new dedispersed signal power would be:
\begin{equation}
  \widehat{P_{dedisp}}[n] = P_{dedisp}[n] + \bar{P}(t_\varsigma+n\,\Delta t,\nu_\varsigma),
\end{equation}
and the new dedispersed noise error would be:
\begin{equation}
  \widehat{\sigma_{n_{dedisp}}} = \sqrt{N_\mathbb{S}+1}\,\frac{k\,T_{sys}\,\Delta\nu}{\sqrt{\Delta\nu\,\Delta t}} = \sqrt{\frac{N_\mathbb{S}+1}{N_\mathbb{S}}}\,\sigma_{n_{dedisp}}.
\end{equation}

The ratio of the new S/N to the old would then be:
\begin{eqnarray}\label{eqn:rel_snr}
  \frac{\widehat{\text{SNR}}[n]}{\text{SNR}[n]} = \frac{\widehat{P_{dedisp}}[n]}{P_{dedisp}[n]}\,\frac{\sigma_{n_{dedisp}}}{\widehat{\sigma_{n_{dedisp}}}}
  = \left(1+\frac{\bar{P}(t_\varsigma+n\,\Delta t,\nu_\varsigma)}{\displaystyle\sum\limits_{s\in\mathbb{S},s\ne\varsigma}\bar{P}(t_s+n\,\Delta t,\nu_s)}\right)\sqrt{\frac{N_\mathbb{S}}{N_\mathbb{S}+1}}.
\end{eqnarray}

On average, we improve the overall S/N by adding sample $\varsigma$ to our sum when eq.~(\ref{eqn:rel_snr}) is greater than unity.  That is, when:
\begin{equation}\label{eqn:sel_criterion}
  \bar{P}(t_\varsigma+n\,\Delta t,\nu_\varsigma) > \left(\sqrt{\frac{N_\mathbb{S}+1}{N_\mathbb{S}}}-1\right)\displaystyle\sum\limits_{s\in\mathbb{S},s\ne\varsigma}\bar{P}(t_s+n\,\Delta t,\nu_s).
\end{equation}

Eq.~(\ref{eqn:sel_criterion}) provides a criterion for adding a new sample ($\varsigma$) to an existing set of samples ($\mathbb{S}$) used in the dedispersion sum for a given trial.  We use this criterion to select, a priori, sets of dynamic spectrum samples to be summed by the dedisperser for each trial.  The $\bar{P}(t,\nu)$ terms, on both sides of the relation, are predicted using eq.~(\ref{Pavg}) and the DM, pulse width and spectral index parameters targeted for the trial.  The discrete time offset, $n$, controls the time at which a dedispersed pulse will appear at the output of the dedisperser relative to the time that the corresponding dispersed pulse arrives at its input, and is therefore chosen to minimize the dedispersion latency and the amount of physical storage required within the dedisperser.

The set of samples that maximizes the S/N may not be unique.  To achieve the maximum S/N with the fewest samples, we recommend the following procedure: Beginning with an empty set, include a sample that has the highest average signal power (as predicted using eq.~(\ref{Pavg})), then add successive samples in order of highest average signal power until eq.~(\ref{eqn:sel_criterion}) is no longer satisfied.

%
%

\section{S/N variation with temporal and spectral resolution} \label{sec:timefreq}

In this section we look at how the signal-to-noise ratio performance of our fast transients detection system varies with temporal and spectral resolution.  We show that systems employing finer resolutions generally achieve better S/N performance than systems employing coarser resolutions, but there is a sweet spot beyond which finer resolutions yield smaller S/N gains.

To simplify the analysis we assume that the scatter broadened pulse has a rectangular profile and that rather than using the procedure described in \S\ref{sec:selection} to select samples, set $\mathbb{S}$ includes any sample that includes a non-zero component of signal power.  That is, $\mathbb{S}$ includes any sample, $s$, for which $\bar{P}(t_s+n\,\Delta t,\nu_s) > 0$.  The signal power for each sample is predicted using eq.~(\ref{Pavg}), the DM, intrinsic pulse width, scatter broadening and spectral index parameters targeted for the trial, and an arbitrary discrete time offset, $n=n_0$.  We have already shown that the S/N can be improved by excluding some samples with small, non-zero amounts of signal power, so the following analysis will use a less than optimal value for the S/N, but this is fine for the purposes of exploring the effects of resolution on the S/N and later in this section we will see how the more rigorous sample selection algorithm improves the S/N.

\begin{figure}[ht]
  \framebox[\textwidth]{\includegraphics[scale=1.5]{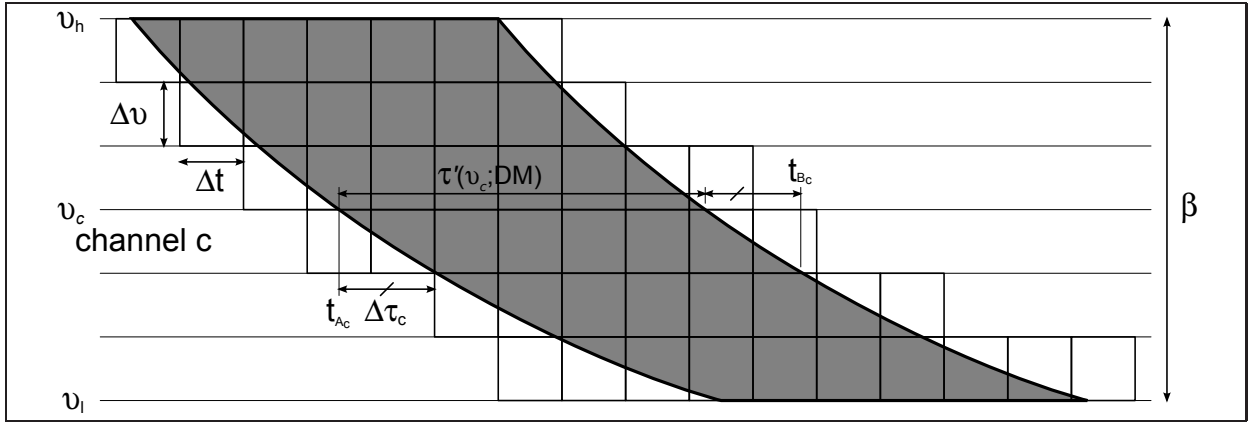}}
  \caption{\label{fig:time_freq_resolution} Sketch of the template dispersed, rectangular pulse (shaded) used to specify the dedispersion set $\mathbb{S}$.  The square cells represent the samples belonging to $\mathbb{S}$.  Each sample includes a non-zero amount of pulse power, and all of the pulse power within bandwidth $\beta$ is collectively included in the samples of $\mathbb{S}$.}
\end{figure}

Figure~\ref{fig:time_freq_resolution} illustrates the profile of the dispersed, rectangular pulse defining the samples of set $\mathbb{S}$.  Here we define $t_{A_c}$ and $t_{B_c}$ to represent the earliest and latest times at which the pulse appears in channel $c$, respectively.  If we define $\nu_c$ to be the highest frequency within channel $c$, then we have $t_{A_c} = \text{DM}/\kappa\,\nu_c^2$, and $t_{B_c} \approx t_{A_c} + \tau'\left(\nu_c;\text{DM}\right) + \Delta\tau_c$.  
Note that the approximation for $t_{B_c}$ assumes that $\tau'\left(\nu_c;\text{DM}\right)$ is approximately constant across the frequency band for channel $c$, which becomes less accurate with coarser spectral resolutions.\footnote{Note that temporal broadening is strongly dependent on frequency (e.g.~$\nu^{-x}$ with $3.5 < x < 4.4$).}
The $\Delta\tau_c$ term is the dispersion smearing time of the signal across channel $c$, which can be approximated as $\Delta\tau_c \approx 2\,\text{DM}\,\Delta\nu/\kappa\,\nu_c^3$.  Thus, if all samples containing non-zero signal power are included in $\mathbb{S}$, then the total number of samples in $\mathbb{S}$ is
\begin{eqnarray}\label{eqn:num_samples}
  N_\mathbb{S} = \displaystyle\sum\limits_{c=0}^{C-1}\left\lceil\frac{t_{B_c}}{\Delta t}\right\rceil - \left\lfloor\frac{t_{A_c}}{\Delta t}\right\rfloor
  \approx \displaystyle\sum\limits_{c=0}^{C-1}\left\lceil\frac{\text{DM}}{\kappa\,\nu_c^2\,\Delta t} + \frac{\tau'\left(\nu_c;\text{DM}\right)}{\Delta t} + \frac{2\,\text{DM}\,\Delta\nu}{\kappa\,\nu_c^3\,\Delta t}\right\rceil - \left\lfloor\frac{\text{DM}}{\kappa\,\nu_c^2\,\Delta t}\right\rfloor .
\end{eqnarray}

If a dispersed pulse matching our prescribed profile is input to this dedisperser, then, by design, the peak S/N will occur at $n=n_0$ and will be
\begin{eqnarray}\label{eqn:snr_approx}
  \text{SNR}[n_0] = \frac{P_0\,\beta\,\tau}{k\,T_{sys}\,\sqrt{\Delta\nu\,\Delta t\,N_\mathbb{S}}}\,\left[\frac{1}{\beta}\,\displaystyle\sum\limits_{c=0}^{C-1}\displaystyle\int_{\nu_c}^{\nu_c + \Delta \nu} d\nu' \left(\frac{\nu'}{\nu_0}\right)^{-\alpha}\right].
\end{eqnarray}

A derivation for eq.~(\ref{eqn:snr_approx}) is given in Appendix~\ref{appendix:SNR_n0}.  It can be shown that for large numbers of frequency channels, i.e.~$C \gg 1$, the term in brackets in eq.~(\ref{eqn:snr_approx}) converges to a constant, and for the remainder of this analysis we assume that this term has little influence on how the peak S/N varies with spectral resolution.  However, note that this assumption does not hold for spectrally steep signals at coarse spectral resolutions, because under these conditions the bracketed term can vary significantly with $C$.

For brevity we define
\begin{eqnarray}\label{eqn:snr_0}
  \text{SNR}_0 = \frac{P_0\,\sqrt{\beta\,\tau}}{k\,T_{sys}}\,\left[\frac{1}{\beta}\,\displaystyle\sum\limits_{c=0}^{C-1}\displaystyle\int_{\nu_c}^{\nu_c + \Delta \nu} d\nu' \left(\frac{\nu'}{\nu_0}\right)^{-\alpha}\right] ,
\end{eqnarray}
which is the maximum S/N that the system would achieve if it had infinitely fine resolution and if there were no scatter broadening.  With scatter broadening, the maximum S/N attenuates with $(\tau/\tau'_{av})^{1/2}$, where $\tau'_{av}$ is the average scatter broadened width of the pulse across the system bandwidth, $\beta$.  Thus, we define the maximum scatter broadened S/N as
\begin{eqnarray}\label{eqn:snr'_0}
  \text{SNR}'_0 = \text{SNR}_0\,\sqrt{\frac{\tau}{\tau'_{av}}} \quad\text{, where}\quad \tau'_{av} = \frac{1}{C}\,\displaystyle\sum\limits_{c=0}^{C-1}\tau'\left(\nu_c;\text{DM}\right) .
\end{eqnarray}

We also define the nominal spectral resolution
\begin{eqnarray}\label{eqn:delta_nu_0}
  \Delta\nu_0 = \frac{\kappa\,\tau'_{av}}{2\,\text{DM}}\,\left[\frac{1}{C}\,\displaystyle\sum\limits_{c=0}^{C-1}\nu_c^{-3}\right]^{-1} ,
\end{eqnarray}
which is the spectral resolution at which the average intra-channel smearing time (i.e.~the average of $\Delta\tau_c$ over all channels) equals the average scatter broadened pulse width, $\tau'_{av}$.  Note that the term in brackets in eq.~(\ref{eqn:delta_nu_0}) is a function of the number of spectral channels, $C$, and converges to a constant for large $C$.

With these definitions, we now examine the effects of spectral and temporal resolution on the peak S/N using the approximations in equations~(\ref{eqn:num_samples}) and (\ref{eqn:snr_approx}).  We consider nine cases, one for each combination of ``fine'', ``nominal'' and ``coarse'' resolution in frequency and time, and for each case the reductions of equations~(\ref{eqn:num_samples}) and (\ref{eqn:snr_approx}) are captured in Table~\ref{tbl:snr_vs_dt_dv}.

\begin{table}[h]
  \caption{Effects of temporal and spectral resolution on S/N.}\label{tbl:snr_vs_dt_dv}
  \begin{longtable}{ | c | c | c | c | }
    \hline                  & \multicolumn{3}{c | }{$\Delta t$} \\
    \cline{2-4} $\Delta\nu$ & {\bf Fine} & {\bf Nominal} & {\bf Coarse} \\
    \hline {\bf Fine}       &
    $\Delta t \ll \tau'_{av}$ &
    $\Delta t \approx \tau'_{av}$
    & $\Delta t \gg \tau'_{av}$ \\
    $\Delta\nu \ll \Delta\nu_0$ &
    $N_\mathbb{S} \approx \frac{\tau'_{av}\,C}{\Delta t}$ &
    $N_\mathbb{S} \approx 2\,C$                     &
    $N_\mathbb{S} \approx C$ \\  &
    $\text{SNR}[n_0] \approx \text{SNR}'_0$ &
    $\text{SNR}[n_0] \approx \frac{\text{SNR}'_0}{\sqrt{2}}$ &
    $\text{SNR}[n_0] \approx \text{SNR}'_0\,\sqrt{\frac{\tau'_{av}}{\Delta t}}$ \\
    \hline {\bf Nominal}    &
    $\Delta t \ll 2\,\tau'_{av}$ &
    $\Delta t \approx 2\,\tau'_{av}$ &
    $\Delta t \gg 2\,\tau'_{av}$ \\
    $\Delta\nu \approx \Delta\nu_0$ &
    $N_\mathbb{S} \approx \frac{2\,\tau'_{av}\,C}{\Delta t}$  &
    $N_\mathbb{S} \approx 2\,C$ &
    $N_\mathbb{S} \approx C$ \\ &
    $\text{SNR}[n_0] \approx \frac{\text{SNR}'_0}{\sqrt{2}}$ &
    $\text{SNR}[n_0] \approx \frac{\text{SNR}'_0}{2}$ &
    $\text{SNR}[n_0] \approx \text{SNR}'_0\,\sqrt{\frac{\tau'_{av}}{\Delta t}}$ \\
    \hline {\bf Coarse}     &
    $\Delta t \ll \frac{\tau'_{av}\,\Delta\nu}{\Delta\nu_0}$ &
    $\Delta t \approx \frac{\tau'_{av}\,\Delta\nu}{\Delta\nu_0}$ &
    $\Delta t \gg \frac{\tau'_{av}\,\Delta\nu}{\Delta\nu_0}$ \\
    $\Delta\nu \gg \Delta\nu_0$  &
    $N_\mathbb{S} \approx \frac{\tau'_{av}\,\beta}{\Delta t\,\Delta\nu_0}$ &
    $N_\mathbb{S} \approx 2\,C$ &  
    $N_\mathbb{S} \approx C$ \\ &
    $\text{SNR}[n_0] \approx \text{SNR}'_0\,\sqrt{\frac{\Delta\nu_0}{\Delta\nu}}$ &
    $\text{SNR}[n_0] \approx \frac{\text{SNR}'_0}{\sqrt{2}}\,\sqrt{\frac{\tau'_{av}}{\Delta t}}$ &
    $\text{SNR}[n_0] \approx \text{SNR}'_0\,\sqrt{\frac{\tau'_{av}}{\Delta t}}$ \\
    & & $\approx \frac{\text{SNR}'_0}{\sqrt{2}}\,\sqrt{\frac{\Delta\nu_0}{\Delta\nu}}$ & \\
    \hline
  \end{longtable}
\end{table}

In Table~\ref{tbl:snr_vs_dt_dv} we show that the peak S/N converges to the optimal value, $\text{SNR}'_0$, at fine temporal and spectral resolutions; drops down to $\text{SNR}'_0/2$ at nominal resolutions; then decays proportional to $\Delta t^{-1/2}$ and $\Delta \nu^{-1/2}$ at coarse resolutions.  Note that while the condition for nominal spectral resolution is the same across all temporal resolutions, the conditions for nominal temporal resolution vary with the spectral resolution, from $\Delta t \approx \tau'_{av}$ at fine spectral resolutions, $\Delta t \approx 2\,\tau'_{av}$ at nominal spectral resolutions, to $\Delta t \approx \tau'_{av}\,\Delta\nu/\Delta\nu_0$ at coarse spectral resolutions.  These conditions are consistent in that each represents the temporal resolution at which the peak S/N is $1/\sqrt{2}$ of the value it would be at an infinitesimally fine temporal resolution.

In the special case where both the temporal resolution and the average intra-channel smearing are limited to some arbitrary multiple of the pulse width, $\Delta t = \text{Av.}[\Delta\tau_c] = m\,\tau'_{av}$, i.e.~$\Delta t/\tau'_{av} = \Delta\nu/\Delta\nu_0 = m$, we have
\begin{eqnarray}
  N_\mathbb{S} \approx \left(2+1/m\right)\,C \quad\text{and}\quad \text{SNR}[n_0] \approx \frac{\text{SNR}'_0}{\sqrt{1+2\,m}} .
\end{eqnarray}
For example, for rectangular pulses, S/Ns greater than 82\% of optimal ($0.82\,\text{SNR}'_0$) can be achieved when the temporal resolution and intra-channel smearing are less than a quarter of the averaged scatter broadened pulse width ($m<0.25$).

\begin{figure}[ht]
  \framebox[\textwidth]{\includegraphics[scale=1.15,trim=0cm -0.3cm 0cm -0.3cm,clip=true]{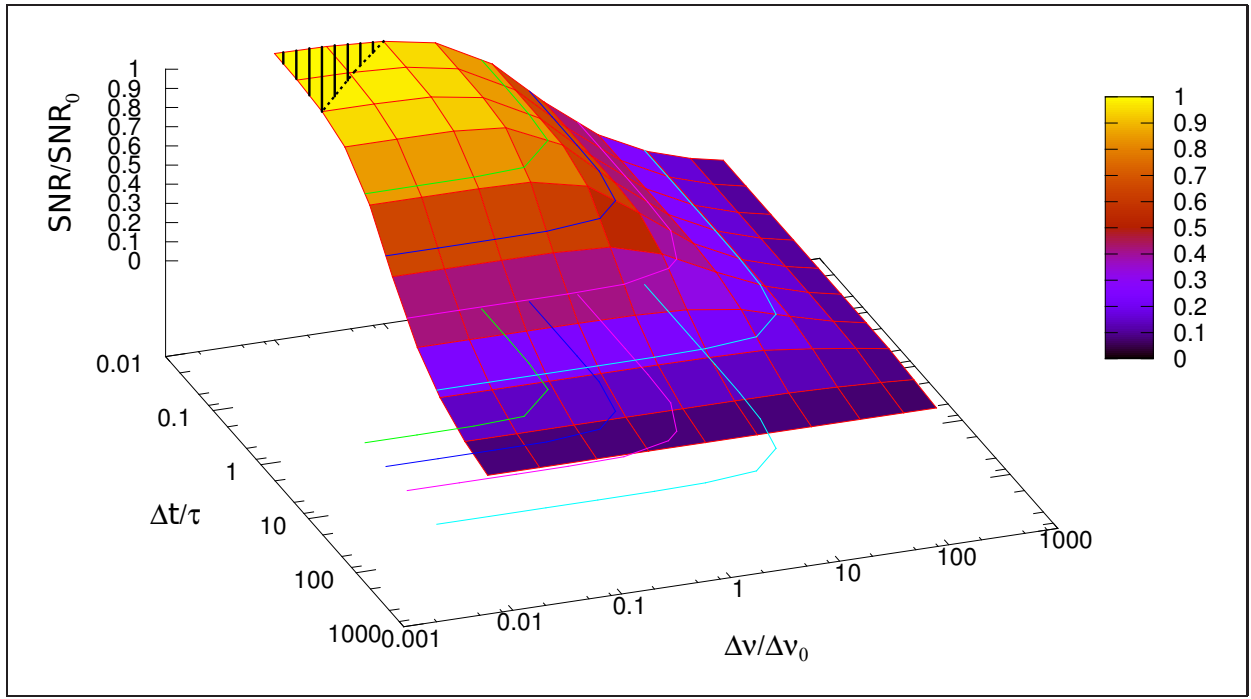}}
  \caption{\label{fig:snr_vs_dt_dv} A surface plot of the dedispersed S/N as a function of temporal and spectral resolution.  The surface points are numerically calculated using the sample selection algorithm given in \S\ref{sec:selection} for a frequency range of 700~MHz to 1~GHz, an intrinsic pulse width of 1~ms, a spectral index of 0, a dispersion measure of 30~pc.cm$^{-3}$ and negligible scatter broadening (such as for an extra-galactic fast transients survey away from the galactic plane).
The hatched region identifies where the temporal-spectral resolution product falls below unity and the analysis is no longer valid.}
\end{figure}

So far we have used a simple and intuitive means of estimating the S/N.  The surface plot in Figure~\ref{fig:snr_vs_dt_dv} justifies this by showing how the peak S/N varies with the spectral and temporal resolution of our system when employing the sample selection algorithm described in \S\ref{sec:selection}.  One sees that the curve conforms with the characteristics described in the above analysis.  The sample selection algorithm slightly improves the S/N predicted in the above analysis, and in this case the relative improvement in S/N is at a maximum of $\sim$20\% at the nominal resolution point.

Table~\ref{tbl:snr_vs_dt_dv} and Figure~\ref{fig:snr_vs_dt_dv} show that significant improvements in S/N performance can be realized by increasing the resolution of a system from coarse to nominal regimes, but also that progressively less S/N improvement can be achieved as the resolution increases beyond nominal.
Finer resolutions generally come at the cost of higher data volumes and faster processing, and these costs need to be weighed against the improvement in performance and overall science goals of the survey.

\section{S/N variation with dispersion measure} \label{sec:snr_vs_dm_approx}

The S/N performance of an incoherent dedispersion system also depends on the dispersion measure of the trial.  This dependence can be seen from the relations in Table~\ref{tbl:snr_vs_dt_dv}.  By rearranging the relations in the left-hand column of Table~\ref{tbl:snr_vs_dt_dv}, the table becomes a description of the S/N for nine regions of DM vs $\Delta t$ space.  Thus, for a given spectral resolution, the DM dimension is divided into ``low'' ($\text{DM} \ll \text{DM}_\tau$), ``nominal'' ($\text{DM} \approx \text{DM}_\tau$)  and ``high'' ($\text{DM} \gg \text{DM}_\tau$) dispersion measures, where
\begin{eqnarray}\label{eqn:DM_tau}
  \text{DM}_\tau = \frac{\tau'_{av}\,\kappa}{2\,\Delta\nu}\,\left[\frac{1}{C}\,\displaystyle\sum\limits_{c=0}^{C-1}\nu_c^{-3}\right]^{-1}.
\end{eqnarray}

Similarly, by replacing the normalized $\Delta\nu$ axis with a normalized DM axis (DM/$\text{DM}_\tau$), the surface plot in Figure~\ref{fig:snr_vs_dt_dv} serves to illustrate the variation in S/N with dispersion measure.

$\text{DM}_\tau$ is the dispersion measure at which the average intra-channel smearing time equals the average width of the scatter broadened pulse.  At DMs larger than $\text{DM}_\tau$, intra-channel smearing losses start to become significant, causing the S/N to decay proportional to $\text{DM}^{-1/2}$.  Intra-channel smearing loss is a common problem for all incoherent dedispersion systems \citep{hankins-sp1975} and such systems often require finer spectral resolutions in order to target larger dispersion measures without significant loss in S/N.

\section{Comparison with matched filters}\label{sec:comparison}

We now compare the performance of the Tardis dedisperser detector to the performance of (i) detectors commonly employed for fast transient and pulsar detection, and (ii) theoretically optimal detectors. The Tardis dedisperser detector operates on power samples in time and frequency space (the dynamic spectrum), whereas conventional fast transient detectors typically operate on time series data, obtained by averaging the dedispersed dynamic spectrum dataset over spectral channels \citep[e.g.,][]{2011ApJ...735...97W,DenCor09}. We consider detectors operating in both the dynamic spectrum (time and frequency samples) and temporal (time samples alone) domains. 

The matched filter detector (MF) is the optimal linear detector for data with generalized Gaussian noise, and operates on the dataset with a replica of the signal profile. When the signal profile is unknown, approximate templates can be evaluated to optimize detection performance. The major disadvantage of the MF is the need for knowledge of the pulse profile. One method of avoiding this issue is to consider a simple boxcar template (specifically, a unit height rectangular pulse with variable width, typically binned geometrically): this detector will have sub-optimal performance compared with the MF for signals that are not rectangular and/or of differing temporal width compared with the template. The boxcar template has been employed in recent fast transients experiments \citep[e.g.,][]{2011ApJ...735...97W,DenCor09}.

In addition to comparing the Tardis dedisperser detector with the time series boxcar detector, we also wish to compare its performance with optimal detectors: the matched filter applied in both the temporal (time series dataset), and dynamic spectrum (time and frequency dataset) domains. The former represents the best-case detector when data are averaged over spectral channels, and the latter represents the best performance achievable with dispersed dynamic spectrum data. Throughout we consider white Gaussian noise, although this is not necessary for the method (knowledge of the noise properties is required, however), and assume that the pulse timing is optimal (pulse arrival aligns with the beginning of a sample, and is known: in general, this is determined empirically). We also omit any contribution to the noise power from signal self-noise, because this is typically small compared with the power contribution from the sky and receiver. This simplification allows us to treat the noise as an additive quantity.

We now use the first and second order statistics of the detection test statistic \citep{kay98} to derive expressions for the signal-to-noise ratios for each of these detectors, and discuss the key differences and similarities between them.

{\bf Time series matched filter (MF):}
Noting that the matched filter multiplies the data by a replica of the expected pulse profile (yielding a square in the following expression), the signal power and noise error are, respectively:
\begin{eqnarray}
P_S &=& \displaystyle\sum\limits_{j=1}^{N_{\text{t}}} \left(\displaystyle\sum\limits_{c=1}^{C} \bar{P}(t_j,\nu_c)\right)^2\\
\sigma_N &=& \frac{kT_{\text{sys}}\sqrt{\Delta\nu}\sqrt{\displaystyle\sum\limits_{j=1}^{N_{\text{t}}} \left(\displaystyle\sum\limits_{c=1}^{C} \bar{P}(t_j,\nu_c)\right)^2}\sqrt{C}}{\sqrt{\Delta{t}}}
\end{eqnarray}
yielding,
\begin{equation}
\text{SNR}_{\text{MF}} = \frac{\sqrt{\displaystyle\sum\limits_{j=1}^{N_{\text{t}}} \left(\displaystyle\sum\limits_{c=1}^{C} \bar{P}(t_j,\nu_c)\right)^2}\sqrt{\Delta{t}}}{kT_{\text{sys}}\sqrt{\Delta\nu}\sqrt{C}},
\end{equation}
where $N_{\text{t}}$ denotes the number of temporal samples, and $C$ is the number of spectral channels.
The number of samples in the MF denominator is $\sim\sqrt{N_{\text{t}}\,C}\gtrsim{\sqrt{N_\mathbb{S}}}$, because the Tardis detector does not have to use all of the spectral channels. This can lead to the MF incorporating more noise power than is optimal if one had the full dynamic dataset (the MF presented here is optimal in the time-series domain). Therefore, the performance of the two detectors depends on the nature of the signal being detected, and consequently, on the spectral and temporal resolution of the experiment.

{\bf Dynamic spectrum matched filter (DSMF):}
\begin{equation}
\text{SNR}_{\text{DSMF}} = \frac{\sqrt{\displaystyle\sum\limits_{j=1}^{N_{\text{t}}} \displaystyle\sum\limits_{c=1}^{C} \bar{P}^2(t_j,\nu_c)}\sqrt{\Delta{t}}}{kT_{\text{sys}}\sqrt{\Delta\nu}}.
\end{equation}
For this detector, all included samples are summed in quadrature. This detector weights each sample according to the expected signal strength, reducing the effective noise contribution to the test statistic. The obvious drawback to implementation is the requirement for full knowledge of the pulse shape.

{\bf Time series boxcar detector (Box.):}
\begin{equation}
\text{SNR}_{\text{BMF}} = \sqrt{\frac{\Delta{t}}{C\,N_{\text{t}}\,\Delta\nu}}\frac{\displaystyle\sum\limits_{j=1}^{N_{\text{t}}} \displaystyle\sum\limits_{c=1}^{C}\bar{P}(t_j,\nu_c)}{kT_{\text{sys}}}.
\end{equation}
This expression is similar to that for the Tardis detector, with the major difference that it is forced to include all of the noise power over the spectral channels. For finite temporal and spectral resolution, and DM$\neq$0, $N_\mathbb{S} < C\,N_{\text{t}}$, and the Tardis detector will always yield improved performance compared with the boxcar MF. Note that this also considers boxcar templates that are optimally matched to the actual signal pulse width: in the general case, when the pulse width is unknown, the boxcar template will not be matched, and the performance will be further degraded.

For a perfectly dedispersed pulse (where $N_\mathbb{S}=C\,N_{\text{t}}$ and $\bar{P}(t,\nu)=\bar{P}$), it is straight-forward to show that all detectors yield the same S/N. It is obvious from these expressions that the Tardis dedisperser detector is a width-optimized boxcar detector in the dynamic spectrum domain (the sample inclusion/exclusion criterion provides the width optimization).

\begin{figure}[ht]
  \framebox[\textwidth]{{\includegraphics[scale=0.45,angle=270]{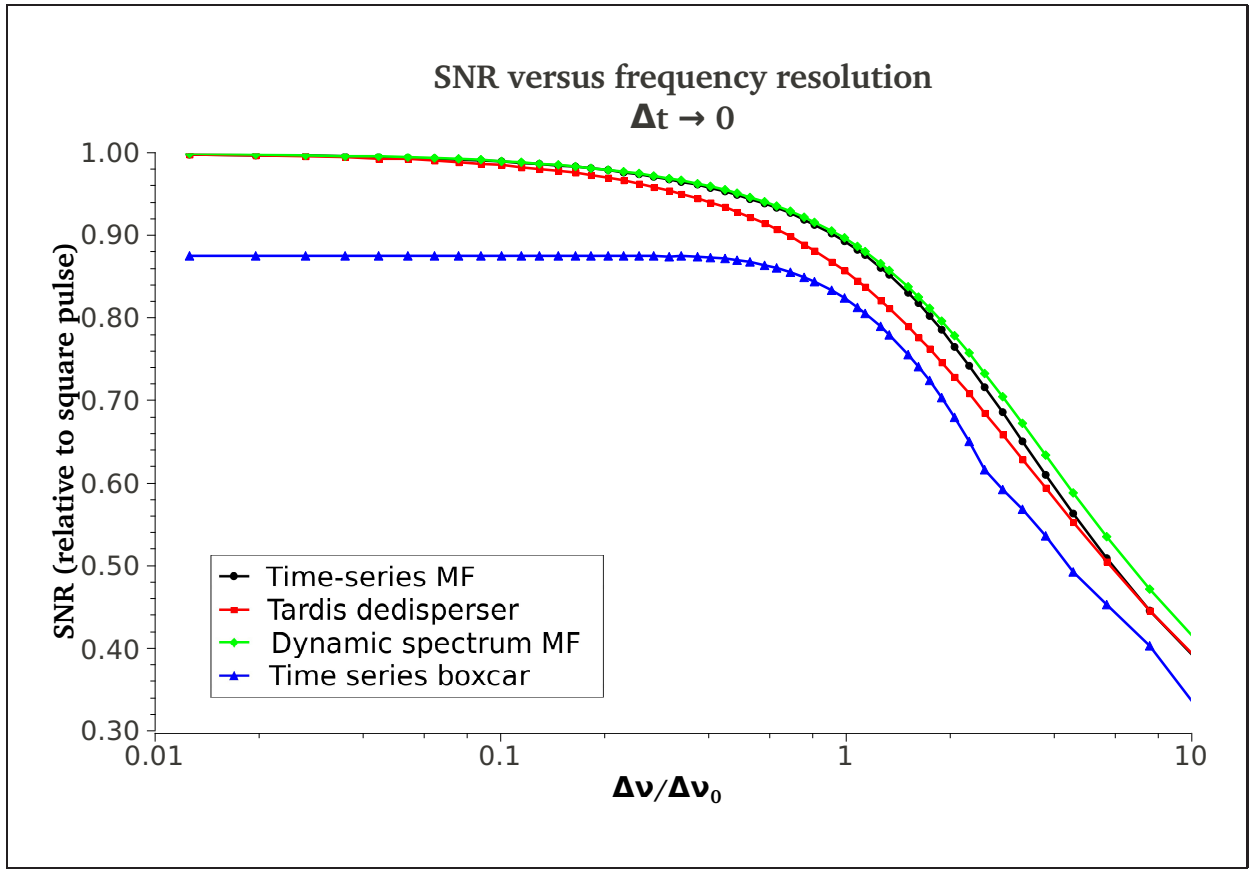}}}
  \caption{\label{fig:snr_detectors} Detection performance signal-to-noise ratios, relative to an un-dispersed rectangular pulse, for the Tardis detector, and three other common detectors.}
\end{figure}

Figure \ref{fig:snr_detectors} displays the detection performance for each detector as a function of the normalized spectral resolution, $\Delta\nu/\Delta\nu_0$. The two matched filter detectors perform well, with the DSMF performing the best across the range tested, as expected. The Tardis dedisperser detector performs well relative to the time series matched filter. At low $\Delta\nu/\Delta\nu_0$ (high resolution), the time series matched filter performs better. This reflects the Tardis detector's binary choice for either including or excluding samples: while excluding a sample may retain a higher S/N, signal power is nonetheless excluded (rather than being optimally-weighted, as for the time series matched filter). At very poor resolution, the two curves cross: the dedispersed signal is substantially broadened at low resolution, and the additional noise power incorporated into the time series matched filter degrades its performance.

The time series boxcar detector has variable performance, depending on how well-matched the coarse temporal bins are to the underlying signal. We have chosen a single possible realization of its performance. Its performance matches that of the others at high resolution, when the pulse width is matched perfectly to a tested bin width, and the start of the pulse aligns with the start of the bin.

The matched filters, for both the time series and dynamic spectrum domains, demonstrate superior performance compared with the Tardis detector, for a wide range of system and signal parameters. Matched filters are, however, difficult to implement in practise, given the need for full knowledge of the signal profile -- a major obstacle for fast transients surveys. On the other hand, boxcar filters, implemented in either the time-series or dynamic spectrum domains, are blind to pulse shape, and suffer performance degradation accordingly (note that the Tardis dedisperser detector is further superior to some time-series implementations [e.g., VFASTR], because it does not use a base-2 discretized temporal binning to produce the trial templates). The Tardis dedisperser detector presented here attempts to balance the performance/signal knowledge trade-off, by exploiting the performance advantages of working in the dynamic spectrum domain and with a sample-selection criterion, to offset performance loss due to lack of pulse shape knowledge. In addition, the Tardis dedisperser detector is computationally efficient to implement, requiring only summing of samples (compared with matched filters, which perform weight and sum operations).

\section{Survey completeness} \label{sec:completeness}

In our analyses so far we have assumed that the dispersion measure of the received signal is known.  However, dispersion measures vary with distance from the source and the content of the intervening ISM along the line-of-sight to the source, and when surveying the sky for new sources, the dispersion measure applicable to each received transient is generally unknown.  It is therefore necessary for the system to dedisperse the signal using a range of trial dispersion measures and search each dedispersed signal for transient content.  Real-time dedispersion and detection processes are compute-intensive, and the computation power increases linearly with the number of trial dispersion measures.  As we will see, the number of trials and the distribution of those trials across the range of dispersion measures targeted by the survey are critical design choices; they determine the completeness of the survey in terms of the average S/N performance.  In this section we describe how a set of trial dispersion measures can be chosen to maximize the completeness of a fast transients survey.

\subsection{Pulse broadening due to DM error} \label{sec:finiteDM}

We begin by illustrating how differences between the dispersion measure assumed for a given trial and the actual dispersion measure of an observed pulse can cause the resulting dedispersed signal to be broadened in time.  Temporal broadening due to DM error is a well documented effect \citep{1969A&A.....2..280B, 2003ApJ...596.1142C, daddario-searching2010} and for completeness we review this in the context of the models presented in this paper.

Assume that the set of samples, $\mathbb{S}$, is chosen to maximize the dedispersed S/N for signals that have a dispersion measure equal to the trial DM, $\text{DM}_\text{trial}$, and assume that we attempt to dedisperse a signal whose actual dispersion measure, $\widehat{\text{DM}}$, differs from the trial DM, i.e.~$\widehat{\text{DM}}\ne \text{DM}_\text{trial}$.  The dedispersed S/N would be:
\begin{eqnarray}\label{eqn:snr_hat}
  \widehat{\text{SNR}}[n] = \sqrt{\frac{\Delta t}{N_\mathbb{S}\,\Delta\nu}}\,\frac{\displaystyle\sum\limits_{s\in\mathbb{S}}{\bar P}(t_s+n\,\Delta t,\nu_s,\widehat{\text{DM}})}{k\,T_{sys}}.
\end{eqnarray}

The ratio of eq.~(\ref{eqn:snr}) and eq.~(\ref{eqn:snr_hat}) gives the ``relative'' S/N for a signal whose DM does not equal the trial DM:
\begin{eqnarray}\label{eqn:rel_snr2}
  \text{SNR}_{rel}[n] = \frac{\widehat{\text{SNR}}[n]}{\text{SNR}[n]} = \frac{\displaystyle\sum\limits_{s\in\mathbb{S}}{\bar P}(t_s+n\,\Delta t,\nu_s,\widehat{\text{DM}})}{\displaystyle\sum\limits_{s\in\mathbb{S}}{\bar P}(t_s+n\,\Delta t,\nu_s,\text{DM}_\text{trial})}.
\end{eqnarray}

\begin{figure}[ht]
  \framebox[\textwidth]{\includegraphics[scale=1.1,angle=270]{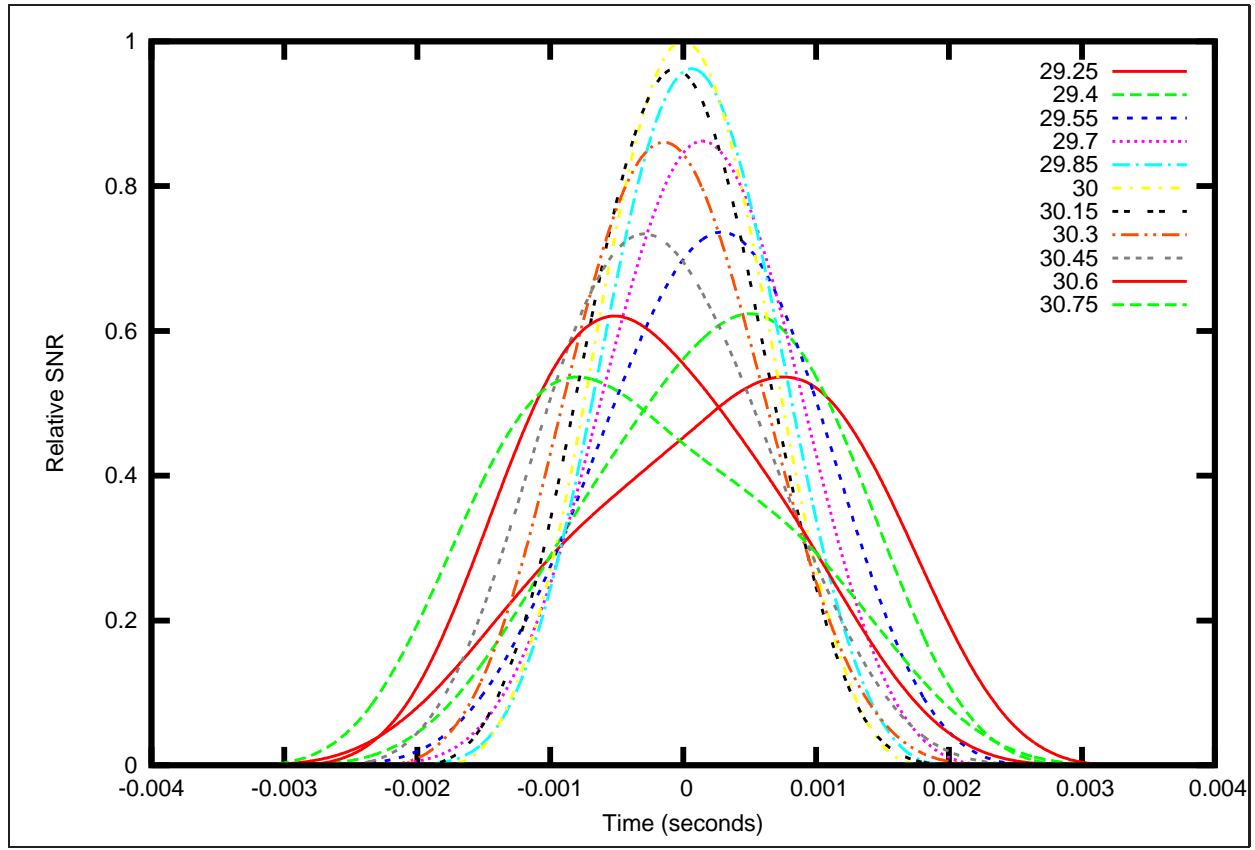}}
  \caption{\label{fig:dm_analysis_askap}Dedispersed pulse profiles for a range of dispersion measures about a trial DM of 30~pc.cm$^{-3}$.  The system bandwidth ranges from 700~MHz to 1004~MHz, with 1~MHz channel resolution and 1~ms temporal resolution.  The scatter broadened profiles of the input pulses (prior to dispersion) are identical: rectangular with width 1~ms.}
\end{figure}

Using the sample selection criterion outlined in \S\ref{sec:selection} for determining $\mathbb{S}$, and the relation for the relative S/N in eq.~(\ref{eqn:rel_snr2}), the profiles for a series of test pulses, each with differing dispersion measures, are plotted in Figure~\ref{fig:dm_analysis_askap} for a trial DM of 30 pc.cm$^{-3}$.  These examples are calculated for signals received in the 700~MHz to 1004~MHz frequency band, with 1~MHz channel resolution and square-law detected samples integrated to a temporal resolution of 1~ms.  Each dispersed pulse input to the dedisperser is modeled using eq.~(\ref{Pavg}) with a rectangular scatter broadened pulse profile of width 1~ms.

The plots show how pulses become increasingly smeared as the differences between the trial and actual DMs increase.  The visible asymmetries in the dedispersed pulses, more notable for those with larger absolute DM errors, are a consequence of the natural $\nu^{-2}$ bend in the dispersion curve.  If the actual DM of the pulse is less than that of the trial, then the trial will initially intersect the dispersed pulse in the low frequency channels, and the point of intersection will progress to the higher frequency channels as the trial sweeps past the dispersed pulse.  Since both the trial and the pulse are more dispersed at lower frequencies, the leading edge of the resulting dedispersed pulse is more extended than its trailing edge.  The converse occurs for pulses with DMs larger than that of the trial: as the trial sweeps past the dispersed pulse, the point of intersection moves from high to low frequency channels, causing the trailing edge of the resulting dedispersed pulse to be more extended than its leading edge.



\subsection{S/N variation with DM error}

The temporal broadening of a pulse due to the difference between its true dispersion measure and a given trial DM (i.e.~the DM error) reduces the S/N of the dedispersed signal for that trial.  \citet{2003ApJ...596.1142C} shows that, in general, temporal broadening reduces the S/N according to
\begin{eqnarray}\label{eqn:snrb/snri}
  \frac{\text{SNR}_b}{\text{SNR}_i} = \sqrt{\frac{W_i}{W_b}},
\end{eqnarray}
where $W_i$ and $\text{SNR}_i$ are the temporal width and S/N of the ``incident'' pulse (i.e.~before the pulse is broadened), and $W_b$ and $\text{SNR}_b$ are the temporal width and S/N of the broadened pulse.  If we consider $W_i$ to be the width of the pulse after it has been dedispersed to a perfectly matched trial DM, such that there is no DM error, then $W_i$ can be approximated using
\begin{eqnarray}
  W_i \approx \sqrt{{\Delta t}^2_{\text{DM}residual} + {\Delta t}^2 + \tau'^2_{av}},
\end{eqnarray}
and if $W_b$ is the width of the pulse after it has been dedispersed to an un-matched trial DM, with a DM error of $\delta\text{DM}$, then
\begin{eqnarray}
  W_b \approx \sqrt{{\Delta t}^2_{\text{DM}residual} + {\Delta t}^2_{\delta\text{DM}} + {\Delta t}^2 + \tau'^2_{av}},
\end{eqnarray}
where $\Delta t_{\text{DM}residual}$ is the component of the pulse width due to residual dispersion smearing (that which cannot be corrected for by dedispersion); ${\Delta t}_{\delta\text{DM}}$ is the component due to the DM error, $\delta\text{DM}$; and as defined earlier, $\Delta t$ and $\tau'_{av}$ are the temporal resolution of our dedispersion system and the average scatter broadened width of the pulse, respectively.

For coherent dedispersion systems, the residual smearing after dedispersion is essentially zero (i.e.~$\Delta t_{\text{DM}residual} = 0$).  However, incoherent dedispersion systems can only remove inter-channel smearing; the residual (intra-)channel smearing can be approximated as
\begin{eqnarray}\label{eqn:residual smearing}
  {\Delta t}_{\text{DM}residual} \approx \frac{2\,\text{DM}\,\Delta\nu}{\kappa\,\nu^3}.
\end{eqnarray}

The component of smearing due to DM error, ${\Delta t}_{\delta\text{DM}}$, is equivalent to the smearing of a signal with a dispersion measure equal to $\delta\text{DM}$ across the entire frequency band, $\beta$.  This smearing can likewise be approximated as
\begin{eqnarray}\label{eqn:dm error smearing}
  {\Delta t}_{\delta\text{DM}} \approx \frac{2\,\delta\text{DM}\,\beta}{\kappa\,\nu^3}.
\end{eqnarray}

By substituting these approximations back into eq.~(\ref{eqn:snrb/snri}) it follows that
\begin{eqnarray}\label{eqn:snrb/snri_approx}
  \frac{\text{SNR}_b}{\text{SNR}_i} \approx \left(\frac{C^2\,\delta\text{DM}^2}{\text{DM}^2 + \text{DM}_{diag}^2 + \text{DM}_\tau^2} + 1\right)^{-1/4},
\end{eqnarray}
where $\text{DM}_{diag}$ is known as the ``diagonal DM'', i.e.~the DM at which the average smearing time across each channel equals the temporal resolution; and $\text{DM}_\tau$ is the DM at which the average smearing time across each channel equals the average width of the scatter broadened pulse.
\begin{eqnarray}
  \text{DM}_{diag} = \frac{\Delta t\,\kappa}{2\,\Delta\nu}\,\left[\frac{1}{C}\,\displaystyle\sum\limits_{c=0}^{C-1}\nu_c^{-3}\right]^{-1} \quad \text{and} \quad \text{DM}_\tau = \frac{\tau'_{av}\,\kappa}{2\,\Delta\nu}\,\left[\frac{1}{C}\,\displaystyle\sum\limits_{c=0}^{C-1}\nu_c^{-3}\right]^{-1}.
\end{eqnarray}

\begin{figure}[ht]
  \framebox[\textwidth]{\includegraphics[scale=0.8]{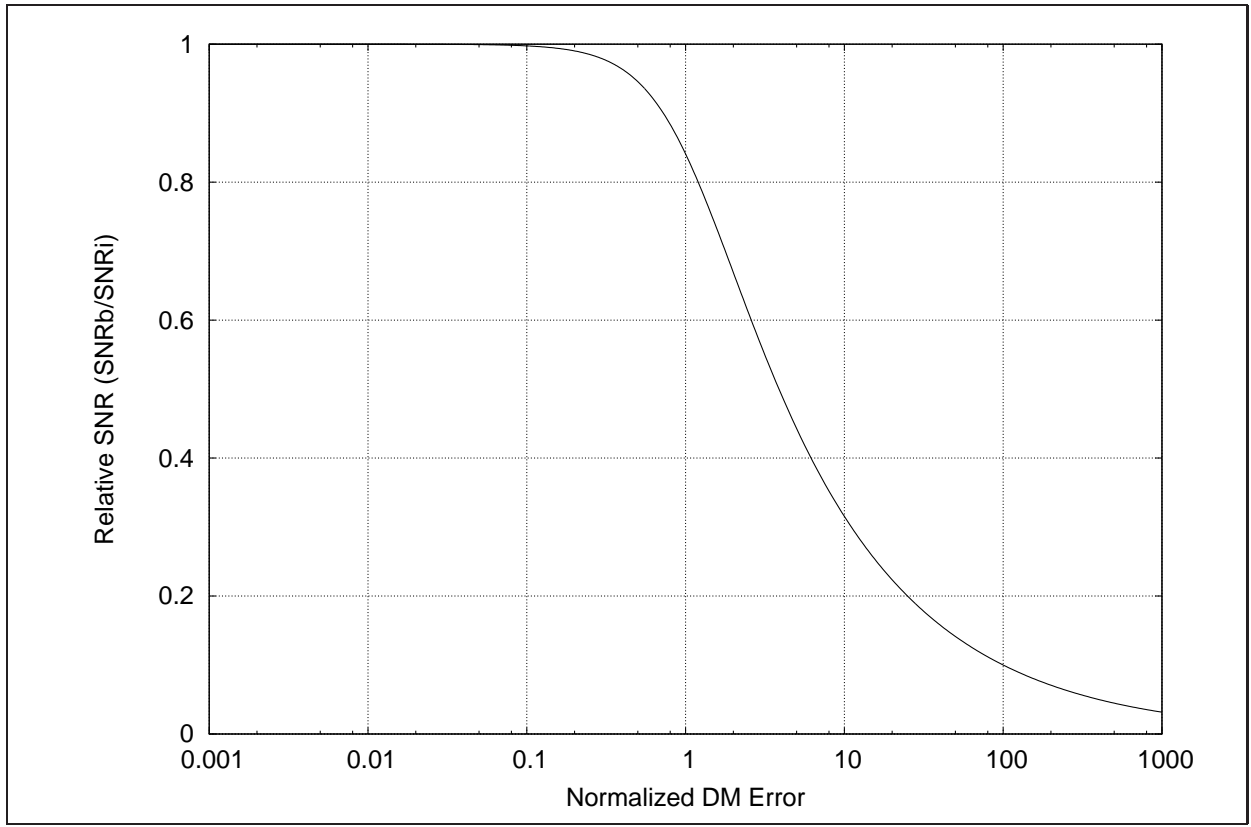}}
  \caption{\label{fig:snr_vs_delta_dm} Attenuation of dedispersed S/N with increasing DM error.  The ordinate is normalised to the dedispersed S/N expected when there is no DM error, i.e.~when the trial DM matches the true DM of the signal.  The abscissa is normalised to $\frac{1}{C}\,\sqrt{\text{DM}^2 + \text{DM}_{diag}^2 + \text{DM}_\tau^2}$.}
\end{figure}

Using the approximation given in eq.~(\ref{eqn:snrb/snri_approx}), Figure~\ref{fig:snr_vs_delta_dm} illustrates how the S/N attenuates as the DM error increases.  The DM error is normalised to a value of $\frac{1}{C}\,\sqrt{\text{DM}^2 + \text{DM}_{diag}^2 + \text{DM}_\tau^2}$, which implies that for sufficiently small dispersion measures, $\text{DM} \ll \text{DM}_{diag}$, or $\text{DM} \ll \text{DM}_\tau$, the S/N attenuation for a given DM error is independent of the dispersion measure; whereas for sufficiently large dispersion measures, $\text{DM} \gg \text{DM}_{diag}$ and $\text{DM} \gg \text{DM}_\tau$, the S/N attenuation for a given DM error is expected to be less for larger dispersion measures.  With greater scatter broadening, the normalisation value increases, which means that although the overall S/N ($\text{SNR}_i$) reduces with scatter broadening, DM errors cause less attenuation in the relative S/N.

\subsection{Choosing trial dispersion measures}

To maximize the average S/N performance of our detection system across all signals within the DM range of our survey, we aim to choose a set of trial DMs that maintains a limited S/N attenuation between trials.  We do so by constraining the relative S/N to some minimum constant value
\begin{eqnarray}\label{eqn:dm error smearing limit}
  \frac{\text{SNR}_b}{\text{SNR}_i} > \left(\epsilon^2 + 1\right)^{-1/4},
\end{eqnarray}
where $\epsilon$ is referred to as the DM error factor.  It can be shown that this constraint is equivalent to limiting the temporal broadening due to DM error to $\Delta t_{\delta\text{DM}} < \epsilon\,W_i$.

By substituting the approximation for the relative S/N given in eq.~(\ref{eqn:snrb/snri_approx}) into eq.~(\ref{eqn:dm error smearing limit}) we can show that the DM error needs to be constrained to
\begin{eqnarray}
  \delta\text{DM} < \frac{\epsilon}{C}\,\sqrt{\text{DM}^2 + \text{DM}_{diag}^2 + \text{DM}_\tau^2}.
\end{eqnarray}

Therefore, given some arbitrary limit to the reduction in S/N that we are prepared to accept between trial DMs (i.e.~$\left(\epsilon^2 + 1\right)^{-1/4}$), and noting that we can space our trial DMs at intervals of $2\,\delta\text{DM}$, we can choose a set of trial dispersion measures that adhere to this limit as follows:
\begin{eqnarray}\label{eqn:trial DMs}
  \text{DM}_n = \text{DM}_0 + \frac{2\,\epsilon}{C}\,\displaystyle\sum\limits_{i=0}^{n-1}\sqrt{\text{DM}_i^2 + \text{DM}_{diag}^2 + \text{DM}_\tau^2},
\end{eqnarray}
where $\text{DM}_i, \forall i\in[0, 1, ..., N-1]$, are the dispersion measures chosen for our set of $N$ trials, with each successive subscript denoting a successively larger dispersion measure.  $\text{DM}_0$ can be set to the minimum dispersion measure in the range to be searched, and each successive trial DM can be calculated from the trial DMs preceeding it using eq.~(\ref{eqn:trial DMs}).  

It follows from eq.~(\ref{eqn:trial DMs}) that where the trial DMs are small, i.e.~where $\text{DM}_n\ll\text{DM}_{diag}$ and $\text{DM}_n\ll\text{DM}_\tau$, and where scatter broadening is either insignificant or independent of the DM, the DM error ($\delta\text{DM}$) is approximately constant and the trial DMs are approximately uniformly (linearly) spaced, i.e.
\begin{eqnarray}
  \text{DM}_n \approx \text{DM}_0 + \frac{2\,\epsilon\,n}{C}\,\sqrt{\text{DM}_{diag}^2 + \text{DM}_\tau^2} .
\end{eqnarray}
But where the trial DMs become dominant (i.e.~$\text{DM}_j\gg\text{DM}_{diag}$ and $\text{DM}_j\gg\text{DM}_\tau$, for some $j<n$), the trial DMs become approximately exponential with $n$, i.e.
\begin{eqnarray}
  \text{DM}_n \approx \text{DM}_j\left(1 + \frac{2\,\epsilon}{C}\right)^{n-j} .
\end{eqnarray}

For coherent dedispersion systems, since there is no channelization and consequently no residual intra-channel smearing, the $\text{DM}_i$ terms disappear from the right-hand side of eq.~(\ref{eqn:trial DMs}) and the trial DMs follow a linear spacing where scatter broadening is small or constant with DM, and become more spread-out at higher DMs where scatter broadening becomes significant.  Therefore generally more trial DMs are needed for coherent dedispersion systems than for incoherent dedispersion systems.

\begin{figure}[ht]
  \framebox[\textwidth]{
    \centering
    \subfigure[Trials distributed using eq.~(\ref{eqn:trial DMs})]{\includegraphics[scale=0.54]{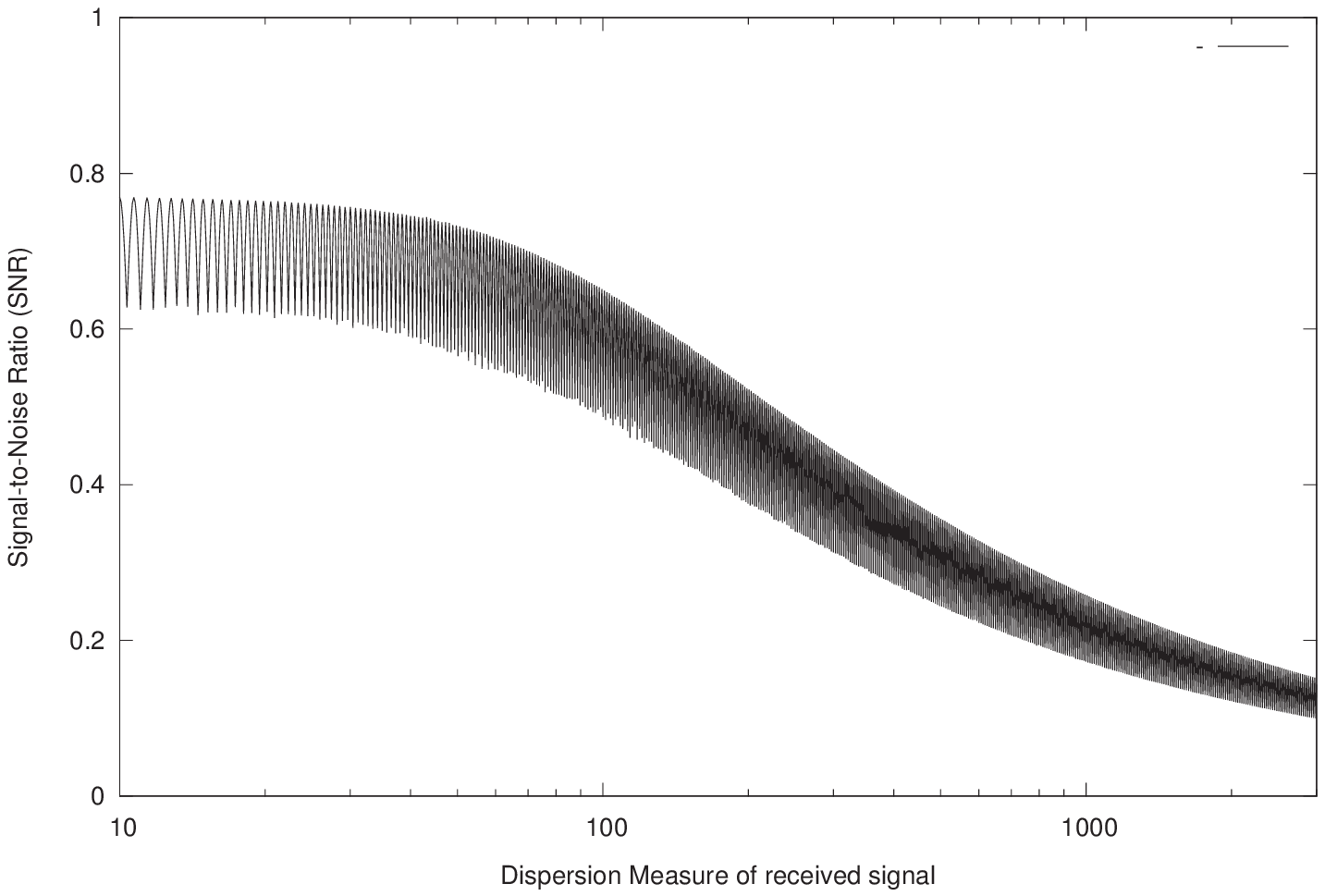}\label{fig:snr_vs_dm_ASKAP_opt}}
    \subfigure[Exponentially distributed trials]{\includegraphics[scale=0.54]{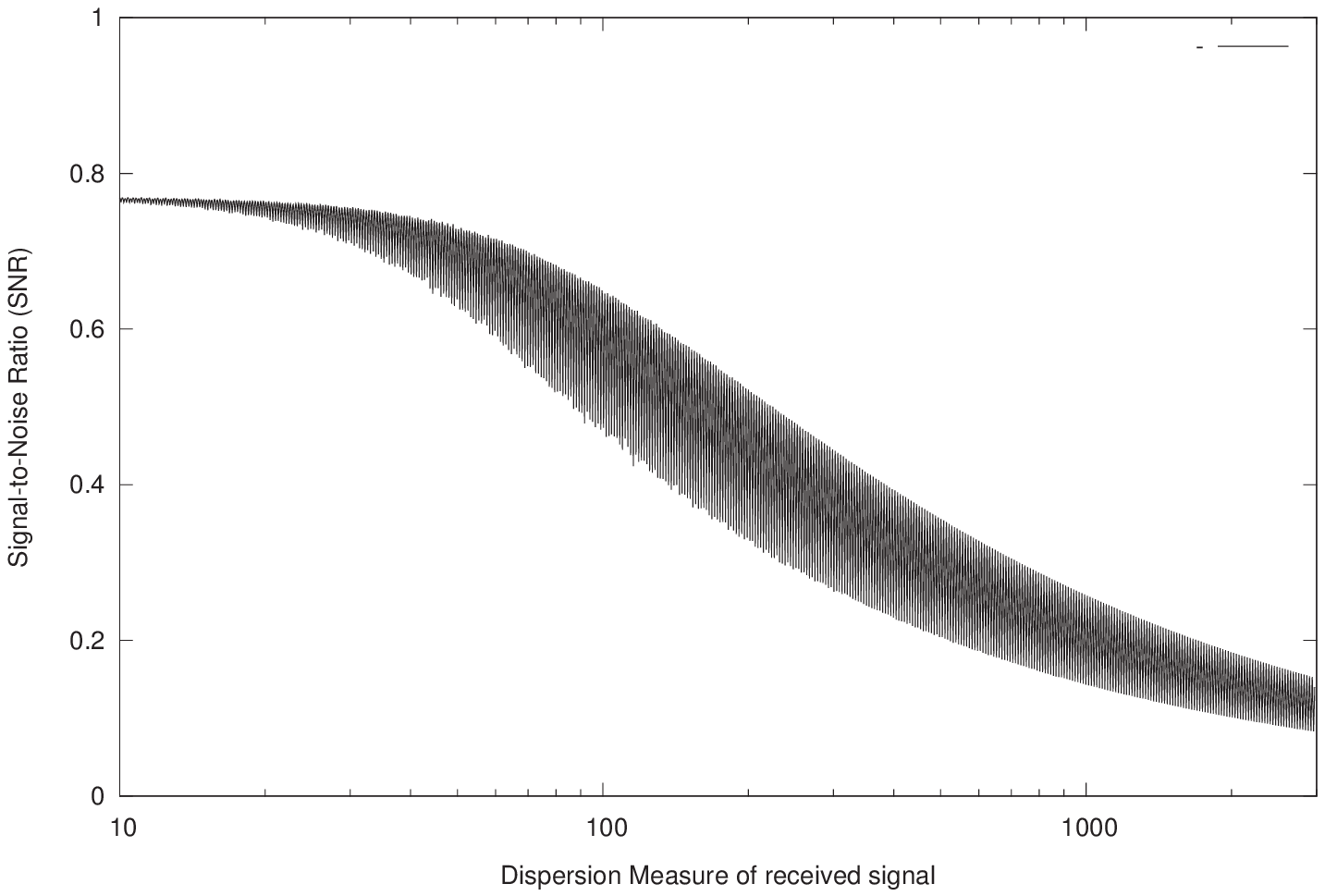}\label{fig:snr_vs_dm_ASKAP_exp}}
  }
  \caption{\label{fig:snr_vs_dm_ASKAP}Plot of the normalised maximum dedispersed S/N as a function of the dispersion measure of a 1~ms test pulse.  In this example, the system bandwidth ranges from 700~MHz to 1~GHz, with 1~MHz channel resolution and 1~ms temporal resolution.  The S/N is the maximum across all trial DMs and normalised to a value of $P_0\,\sqrt{\beta\,\tau}/k\,T_{sys}$.  A total of 442 trial DMs are distributed from 10 to 3000~pc.cm$^{-3}$ using: (a) the relation given in eq.~(\ref{eqn:trial DMs}), with a DM error factor of $\epsilon = 1.492$; and (b) exponentially distributed trials, with a trial ratio of 0.013.}
\end{figure}

Figure~\ref{fig:snr_vs_dm_ASKAP} demonstrates how the choice of trial DMs can impact the S/N performance for the Tardis fast transients detection system planned for the CRAFT survey.  CRAFT aims to survey the sky for milli-second-scale transients from both galactic and extra-galactic sources by making use of ASKAP's wide (30 degree$^2$) field of view.  Given the high luminosities of recently detected extra-galactic fast transients \citep[e.g.,][]{2007Sci...318..777L, 2011MNRAS.415.3065K}, it is reasonable to expect that Tardis may detect sources to redshifts of $z \lesssim 3$, implying IGM dominated dispersion measures up to $\sim$3000~pc.cm$^{-3}$ \citep{2004MNRAS.348..999I}.  Thus a range from 10 to 3000~pc.cm$^{-3}$ is targeted for CRAFT, which the Tardis system intends to cover with 442 trials.

The plot in Figure~\ref{fig:snr_vs_dm_ASKAP_opt} shows the S/N performance where those 442 trials are distributed using the relation given in eq.~(\ref{eqn:trial DMs}).  The S/N consists of a series of finely spaced peaks and troughs, where at the peaks, the true dispersion measure of the pulse matches a trial DM, and at the troughs, the true dispersion measure falls in the middle of two adjacent trial DMs.  
Note the relative drop-out, i.e.~the ratio of the S/N of a trough to the S/N of its adjacent peaks, is constant across the full range of dispersion measures.
The DM error factor in this case is 1.492, giving a S/N drop-out between trial DMs of about 0.75 relative to surrounding peaks.  Also note the general $\text{DM}^{-1/2}$ attenuation in S/N discussed above in \S\ref{sec:snr_vs_dm_approx}.

We compare the plot in Figure~\ref{fig:snr_vs_dm_ASKAP_opt} with the plot in Figure~\ref{fig:snr_vs_dm_ASKAP_exp} where the same number of trials are exponentially distributed across the DM range.  Here we see that the exponential distribution packs trials unnecessarily tightly at low dispersion measures, leaving fewer trials available for higher DMs and overall poorer average performance across the entire range.

\begin{figure}[ht]
  \framebox[\textwidth]{
    \centering
    \subfigure[256 channels]{\includegraphics[scale=0.54]{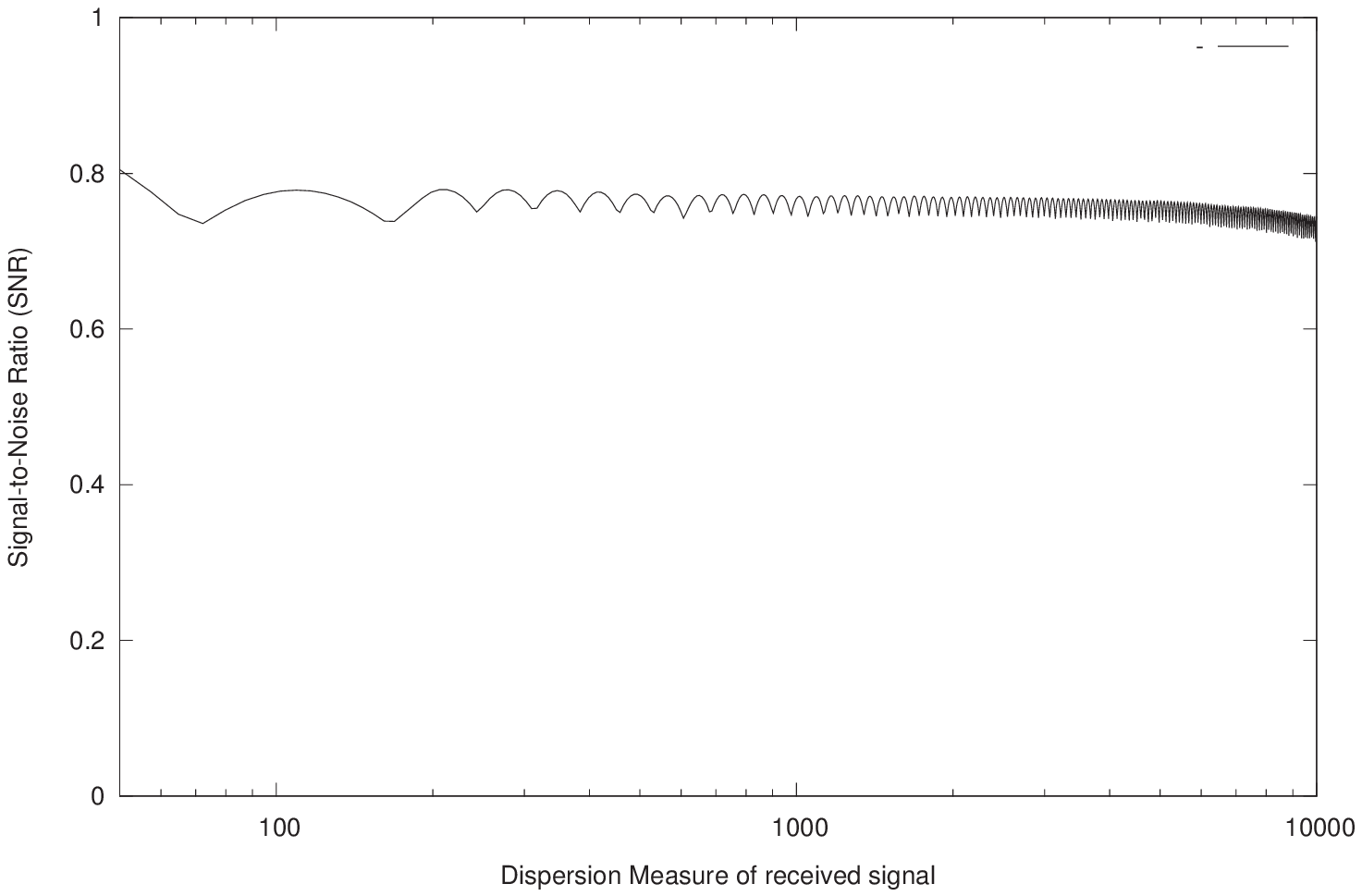}}
    \subfigure[32 channels]{\includegraphics[scale=0.54]{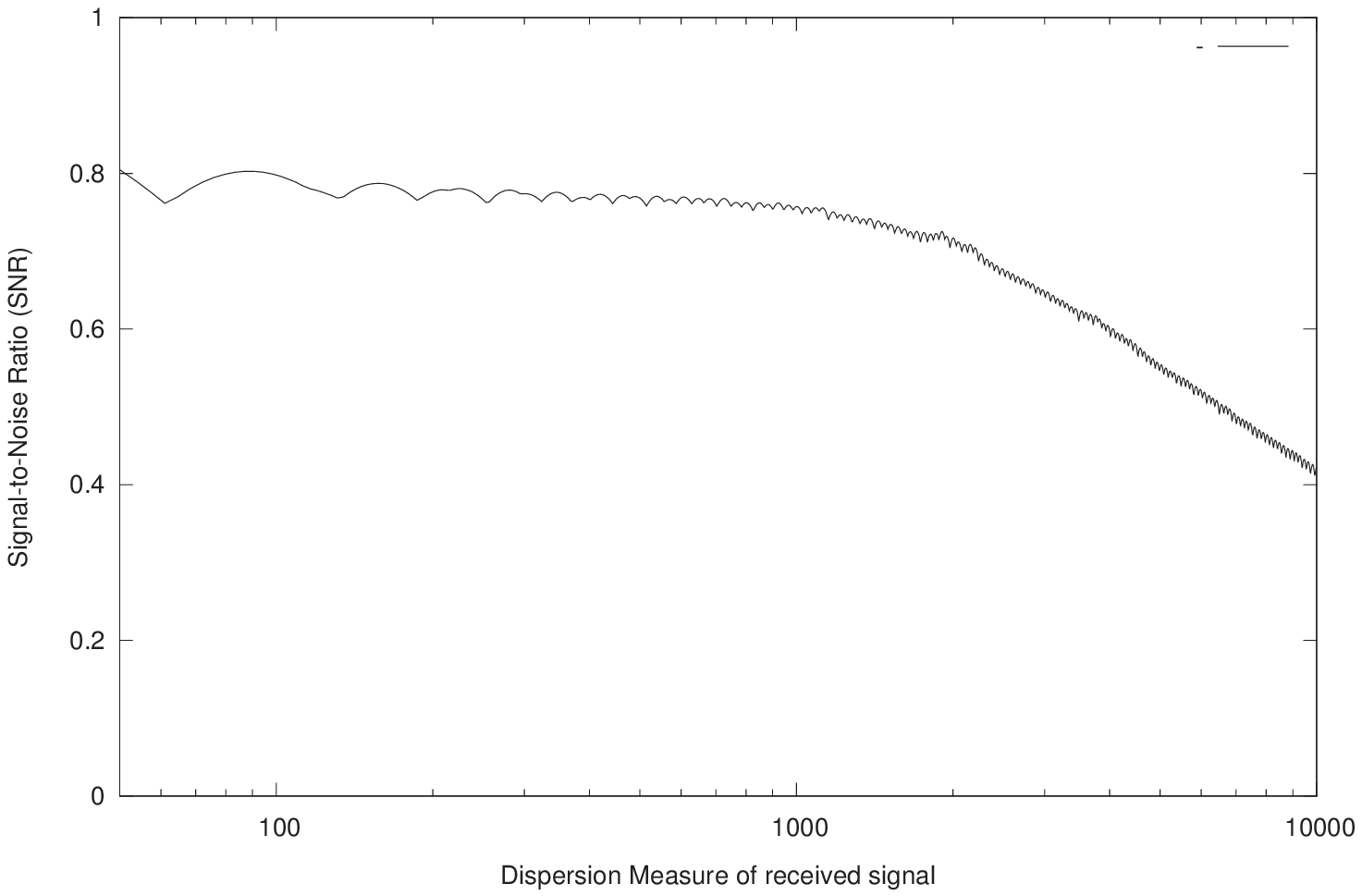}}
  }
  \caption{\label{fig:snr_vs_dm_DSN-43} Plots of the normalised maximum dedispersed S/N as a function of the dispersion measure of a 100~$\mu$s test pulse.  In both plots, the system bandwidth ranges from 21~GHz to 23~GHz, with 100~$\mu$s temporal resolution and spectral resolutions of (a) 7.8125~MHz (256 channels), and (b) 62.5~MHz (32 channels).  The S/N is the maximum across all trial DMs and normalised to a value of $P_0\,\sqrt{\beta\,\tau}/k\,T_{sys}$.  In both cases, a total of 128 trial DMs are distributed from 50 to 10,000~pc.cm$^{-3}$ using the relation given in eq.~(\ref{eqn:trial DMs}) and with DM error factors of (a) $\epsilon = 0.580$, and (b) $\epsilon = 0.287$.}
\end{figure}

It is well known that dedispersion systems with larger numbers of channels (i.e.~finer spectral resolutions) require more trial DMs to achieve the same S/N drop-out between trials \citep{hankins-sp1975}, and this can be seen from the dependence on $C$ in eq.~(\ref{eqn:trial DMs}).  Essentially, systems with coarser spectral resolutions suffer more significant intra-channel smearing, making them less sensitive to DM error than systems with finer spectral resolutions.  This is demonstrated in Figure~\ref{fig:snr_vs_dm_DSN-43} for a putative high radio frequency (21 to 23~GHz) survey for milli-second pulsars at the Galactic Center where dispersion measures as high as $\sim$2000-5000~pc.cm$^{-3}$ can be expected.  For the purposes of this example we consider DMs in the range 50 to 10,000~pc.cm$^{-3}$.  Two possibilities are plotted: In (a), the band is divided into 256 channels of 7.8125~MHz, and in (b), the band is divided into 32 channels of 62.5~MHz.  Each target the same number of trial DMs (128), but to do so, the finer spectral resolution example must suffer a higher DM error factor.  This can be seen in the plots as slightly larger drop-outs between trial DMs: In (a), troughs are 93\% of the peaks; while in (b), troughs are 98\% of the peaks.  The underlying cause of this is that at high DMs the system with coarser spectral resolution (b) suffers from significant intra-channel smearing, making it less sensitive to DM error than the finer spectral resolution system.  Consequently, for the plot in (b), the trial DMs can be more spread-out at high DMs, and more compact at low DMs, resulting in shallower troughs between trial DMs.  On the other hand, the finer spectral resolution system suffers less S/N degradation due to intra-channel smearing and is therefore able to maintain higher overall S/N performance at high dispersion measures.  In terms of survey completeness, the finer spectral resolution example is preferable since it has a higher average S/N over the DM range.

\newpage
\section{Conclusions} \label{sec:conc}

In this paper we have examined the signal-to-noise performance of a new incoherent dedispersion algorithm that improves on the performance of traditional algorithms by supporting multiple temporal bins per spectral channel in the sum that forms the dedispersed time series for a given trial.  
The algorithm has the freedom to include (or exclude) any sample of the dynamic spectrum in its dedispersion sum, thus providing a crude mechanism for matching the profile of a pulse without the computational expense of weighting each sample. 
Even without sample weights, the new algorithm displays comparable S/N performance to the ideal matched filter (both time-domain and dynamic spectrum matched filters) and improved performance over traditional time-series boxcar filters.
Critical parameters affecting S/N performance include the system temperature, frequency range, and spectral and temporal resolutions of the system, the ranges of pulse widths and dispersion measures targeted for the survey, and the number and distribution of trial dispersion measures across the DM range.
Given an assumed pulse profile and dispersion measure for a trial, application of the sample selection criterion presented in this paper ensures that the S/N of the dedispersed time series is optimized with a minimal number of samples.
The paper has demonstrated that significant improvements in S/N performance can be achieved for moderate increases in resolution when both the temporal resolution and the average intra-channel smearing time are approximately equal to the target pulse width.
Progressively less S/N improvement can be achieved as the resolution increases beyond this nominal resolution point, and at coarser resolutions the S/N diminishes with $\Delta t^{-1/2}$ and $\Delta\nu^{-1/2}$.
Once a suitable system resolution has been identified, the number and distribution of trial dispersion measures can be determined, and the paper has presented a new trial DM selection algorithm designed to maintain a predefined minimum relative S/N performance across the targeted range of DMs.

\acknowledgments

The authors are grateful to Peter~Hall, Larry~D'Addario and Stephen Ord for their many and various comments and suggestions.
The Centre for All-sky Astrophysics is an Australian Research Council Centre of Excellence, funded by grant CE110001020. The International Centre for 
Radio Astronomy Research (ICRAR) is a Joint Venture between Curtin University and the University of Western Australia, funded by the State Government of Western Australia and the Joint Venture partners.

\newpage
\appendix

\section{Peak dedispersed S/N for rectangular pulses}\label{appendix:SNR_n0}

We derive an expression for the peak dedispersed signal-to-noise ratio for pulses that have rectangular scatter broadened pulse profiles:
\begin{eqnarray}\label{eqn:profile_sb}
f'(t;\nu,\text{DM}) = f * h_d = \left[\text{H}(t) - \text{H}(t - \tau'(\nu;\text{DM}))\right]\,\frac{\tau}{\tau'(\nu;\text{DM})},
\end{eqnarray}
where $f$ is the intrinsic pulse profile, $h_d$ is the impulse response function for scatter broadening, $\text{H}(t)$ is the Heaviside function, $\tau$ is the intrinsic width of the pulse, and $\tau'(\nu;\text{DM})$ is the pulse width after scatter broadening.  The fractional term on the right of eq.~(\ref{eqn:profile_sb}) accounts for proportional attenuation of the pulse intensity with scatter broadening.  For this profile, the average power of the dispersed pulse in the dynamic spectrum bounded by the temporal and spectral limits $[t,t+\Delta t]$ and $[\nu,\nu+\Delta \nu]$ is
{\scriptsize
\begin{eqnarray}\label{eqn:dispersed pulse}
\bar{P}(t,\nu) &=& \frac{P_0}{\Delta t} \displaystyle\int_{\nu}^{\nu + \Delta \nu} d\nu' \left( \frac{\nu'}{\nu_0} \right)^{-\alpha} \displaystyle\int_t^{t+\Delta t} dt'
\left[ \text{H}\left(t' - \frac{\text{DM}}{\kappa \nu'^2} \right) - \text{H} \left(t'- \tau' - \frac{\text{DM}}{\kappa \nu'^2}\right) \right]\,\frac{\tau}{\tau'} . 
\end{eqnarray}
}

Substituting eq.~(\ref{eqn:dispersed pulse}) into eq.~(\ref{eqn:snr}) gives an expression for the S/N of the $n^\text{th}$ sample of the dedispersed time series:
{\scriptsize
  \begin{eqnarray}\label{eqn:dispersed snr}
    \text{SNR}[n] &=& \frac{P_0}{k\,T_{sys}\,\sqrt{\Delta\nu\,\Delta t\,N_\mathbb{S}}} \displaystyle\sum\limits_{s\in\mathbb{S}} \displaystyle\int_{\nu_s}^{\nu_s + \Delta \nu} d\nu' \left( \frac{\nu'}{\nu_0} \right)^{-\alpha} \displaystyle\int_{t_s+n\,\Delta t}^{t_s+(n+1)\,\Delta t} dt' \left[ \text{H}\left(t' - \frac{\text{DM}}{\kappa \nu'^2} \right) - \text{H} \left(t'- \tau' - \frac{\text{DM}}{\kappa \nu'^2}\right) \right]\,\frac{\tau}{\tau'} \nonumber \\
    &=& \frac{P_0\,\tau}{k\,T_{sys}\,\sqrt{\Delta\nu\,\Delta t\,N_\mathbb{S}}} \displaystyle\sum\limits_{c=0}^{C-1} \displaystyle\int_{\nu_c}^{\nu_c + \Delta \nu} d\nu' \left( \frac{\nu'}{\nu_0} \right)^{-\alpha} \frac{1}{\tau'} \displaystyle\sum\limits_{s\in\mathbb{S},\nu_s=\nu_c} \displaystyle\int_{t_s+n\,\Delta t}^{t_s+(n+1)\,\Delta t} dt' \left[ \text{H}\left(t' - \frac{\text{DM}}{\kappa \nu'^2} \right) - \text{H} \left(t'- \tau' - \frac{\text{DM}}{\kappa \nu'^2}\right) \right] , \nonumber \\
\end{eqnarray}
}
where $\mathbb{S}$ is the set of $N_\mathbb{S}$ dynamic spectrum samples chosen for the dedispersion sum.  If at $n = n_0$ the dedisperser receives a pulse whose profile precisely matches the assumed scatter broadened pulse profile modelled in eq.~(\ref{eqn:profile_sb}), then the samples of set $\mathbb{S}$ will collectively include all of the pulse power and the sum on the right of eq.~(\ref{eqn:dispersed snr}) will equate to the scatter broadened pulse width, $\tau'$, leaving
  \begin{eqnarray}
    \text{SNR}[n_0] &=& \frac{P_0\,\tau}{k\,T_{sys}\,\sqrt{\Delta\nu\,\Delta t\,N_\mathbb{S}}} \displaystyle\sum\limits_{c=0}^{C-1} \displaystyle\int_{\nu_c}^{\nu_c + \Delta \nu} d\nu' \left( \frac{\nu'}{\nu_0} \right)^{-\alpha} \nonumber \\
    &=& \frac{P_0\,\beta\,\tau}{k\,T_{sys}\,\sqrt{\Delta\nu\,\Delta t\,N_\mathbb{S}}} \left[ \frac{1}{\beta} \displaystyle\sum\limits_{c=0}^{C-1} \displaystyle\int_{\nu_c}^{\nu_c + \Delta \nu} d\nu' \left( \frac{\nu'}{\nu_0} \right)^{-\alpha} \right] 
\end{eqnarray}

Note that the term in brackets converges to a constant for increasing numbers of channels ($C$), and for suitably small channel bandwidths ($\Delta\nu$), can be approximated by
  \begin{eqnarray}
    \frac{1}{\beta} \displaystyle\sum\limits_{c=0}^{C-1} \displaystyle\int_{\nu_c}^{\nu_c + \Delta \nu} d\nu' \left( \frac{\nu'}{\nu_0} \right)^{-\alpha} \approx \frac{1}{C} \displaystyle\sum\limits_{c=0}^{C-1} \left( \frac{\nu_c}{\nu_0} \right)^{-\alpha} 
\end{eqnarray}

\newpage
\bibliographystyle{apj}
\bibliography{apj-jour,papers}

\begin{thebibliography}{20}
\expandafter\ifx\csname natexlab\endcsname\relax\def\natexlab#1{#1}\fi

\bibitem[{{Burke-Spolaor} {et~al.}(2011){Burke-Spolaor}, {Bailes}, {Johnston},
  {Bates}, {Bhat}, {Burgay}, {D'Amico}, {Jameson}, {Keith}, {Kramer}, {Levin},
  {Milia}, {Possenti}, {Stappers}, \& {van Straten}}]{burke_spolaor2011}
{Burke-Spolaor}, S., {Bailes}, M., {Johnston}, S., {et~al.} 2011, \mnras, 416,
  2465

\bibitem[{{Burns} \& {Clark}(1969)}]{1969A&A.....2..280B}
{Burns}, W.~R., \& {Clark}, B.~G. 1969, \aap, 2, 280

\bibitem[{{Cordes} \& {McLaughlin}(2003)}]{2003ApJ...596.1142C}
{Cordes}, J., \& {McLaughlin}, M. 2003, \apj, 596, 1142

\bibitem[{{Cordes} {et~al.}(2006){Cordes}, {Freire}, {Lorimer}, {Camilo},
  {Champion}, {Nice}, {Ramachandran}, {Hessels}, {Vlemmings}, {van Leeuwen},
  {Ransom}, {Bhat}, {Arzoumanian}, {McLaughlin}, {Kaspi}, {Kasian}, {Deneva},
  {Reid}, {Chatterjee}, {Han}, {Backer}, {Stairs}, {Deshpande}, \&
  {Faucher-Gigu{\`e}re}}]{cordes2006}
{Cordes}, J., {Freire}, P., {Lorimer}, D., {et~al.} 2006, \apj, 637, 446

\bibitem[{D'Addario(2010)}]{daddario-searching2010}
D'Addario, L. 2010, Searching for Dispersed Transient Pulses with ASKAP, SKA
  Memo 124

\bibitem[{{Deneva} {et~al.}(2009){Deneva}, {Cordes}, {McLaughlin}, {Nice},
  {Lorimer}, {Crawford}, {Bhat}, {Camilo}, {Champion}, {Freire}, {Edel},
  {Kondratiev}, {Hessels}, {Jenet}, {Kasian}, {Kaspi}, {Kramer}, {Lazarus},
  {Ransom}, {Stairs}, {Stappers}, {van Leeuwen}, {Brazier}, {Venkataraman},
  {Zollweg}, \& {Bogdanov}}]{DenCor09}
{Deneva}, J.~S., {Cordes}, J.~M., {McLaughlin}, M.~A., {et~al.} 2009, \apj,
  703, 2259

\bibitem[{Hankins \& Rickett(1975)}]{hankins-sp1975}
Hankins, T.~H., \& Rickett, B.~J. 1975, in Methods in Computational Physics,
  ed. B.~Alder, S.~Fernbach, \& M.~Rotenberg, Vol.~14, {55--129}

\bibitem[{{Inoue}(2004)}]{2004MNRAS.348..999I}
{Inoue}, S. 2004, \mnras, 348, 999

\bibitem[{{Kay}(1998)}]{kay98}
{Kay}, S. 1998, Fundamentals of statistical signal processing: detection theory
  (Prentice-Hall)

\bibitem[{{Keane} {et~al.}(2011){Keane}, {Kramer}, {Lyne}, {Stappers}, \&
  {McLaughlin}}]{2011MNRAS.415.3065K}
{Keane}, E.~F., {Kramer}, M., {Lyne}, A.~G., {Stappers}, B.~W., \&
  {McLaughlin}, M.~A. 2011, \mnras, 415, 3065

\bibitem[{{Keith} {et~al.}(2010){Keith}, {Jameson}, {van Straten}, {Bailes},
  {Johnston}, {Kramer}, {Possenti}, {Bates}, {Bhat}, {Burgay}, {Burke-Spolaor},
  {D'Amico}, {Levin}, {McMahon}, {Milia}, \& {Stappers}}]{keith2010}
{Keith}, M., {Jameson}, A., {van Straten}, W., {et~al.} 2010, \mnras, 409, 619

\bibitem[{{Lorimer} {et~al.}(2007){Lorimer}, {Bailes}, {McLaughlin},
  {Narkevic}, \& {Crawford}}]{2007Sci...318..777L}
{Lorimer}, D.~R., {Bailes}, M., {McLaughlin}, M.~A., {Narkevic}, D.~J., \&
  {Crawford}, F. 2007, Science, 318, 777

\bibitem[{{Macquart} {et~al.}(2010){Macquart}, {Bailes}, {Bhat}, {Bower},
  {Bunton}, {Chatterjee}, {Colegate}, {Cordes}, {D'Addario}, {Deller},
  {Dodson}, {Fender}, {Haines}, {Halll}, {Harris}, {Hotan}, {Jonston}, {Jones},
  {Keith}, {Koay}, {Lazio}, {Majid}, {Murphy}, {Navarro}, {Phillips}, {Quinn},
  {Preston}, {Stansby}, {Stairs}, {Stappers}, {Staveley-Smith}, {Tingay},
  {Thompson}, {van Straten}, {Wagstaff}, {Warren}, {Wayth}, {Wen}, \& {CRAFT
  Collaboration}}]{2010PASA...27..272M}
{Macquart}, J., {Bailes}, M., {Bhat}, N.~D.~R., {et~al.} 2010, \pasa, 27, 272

\bibitem[{{Macquart}(2011)}]{macquart2011}
{Macquart}, J.-P. 2011, \apj, 734, 20

\bibitem[{{Manchester} {et~al.}(2001){Manchester}, {Lyne}, {Camilo}, {Bell},
  {Kaspi}, {D'Amico}, {McKay}, {Crawford}, {Stairs}, {Possenti}, {Kramer}, \&
  {Sheppard}}]{2001MNRAS.328...17M}
{Manchester}, R., {Lyne}, A., {Camilo}, F., {et~al.} 2001, \mnras, 328, 17

\bibitem[{{Rickett}(1990)}]{1990ARA&A..28..561R}
{Rickett}, B.~J. 1990, \araa, 28, 561

\bibitem[{{Stappers} {et~al.}(2011){Stappers}, {Hessels}, {Alexov}, {Anderson},
  {Coenen}, {Hassall}, {Karastergiou}, {Kondratiev}, {Kramer}, {van Leeuwen},
  {Mol}, {Noutsos}, {Romein}, {Weltevrede}, {Fender}, {Wijers}, {B{\"a}hren},
  {Bell}, {Broderick}, {Daw}, {Dhillon}, {Eisl{\"o}ffel}, {Falcke},
  {Griessmeier}, {Law}, {Markoff}, {Miller-Jones}, {Scheers}, {Spreeuw},
  {Swinbank}, {Ter Veen}, {Wise}, {Wucknitz}, {Zarka}, {Anderson}, {Asgekar},
  {Avruch}, {Beck}, {Bennema}, {Bentum}, {Best}, {Bregman}, {Brentjens}, {van
  de Brink}, {Broekema}, {Brouw}, {Br{\"u}ggen}, {de Bruyn}, {Butcher},
  {Ciardi}, {Conway}, {Dettmar}, {van Duin}, {van Enst}, {Garrett}, {Gerbers},
  {Grit}, {Gunst}, {van Haarlem}, {Hamaker}, {Heald}, {Hoeft}, {Holties},
  {Horneffer}, {Koopmans}, {Kuper}, {Loose}, {Maat}, {McKay-Bukowski},
  {McKean}, {Miley}, {Morganti}, {Nijboer}, {Noordam}, {Norden}, {Olofsson},
  {Pandey-Pommier}, {Polatidis}, {Reich}, {R{\"o}ttgering}, {Schoenmakers},
  {Sluman}, {Smirnov}, {Steinmetz}, {Sterks}, {Tagger}, {Tang}, {Vermeulen},
  {Vermaas}, {Vogt}, {de Vos}, {Wijnholds}, {Yatawatta}, \&
  {Zensus}}]{2011A&A...530A..80S}
{Stappers}, B.~W., {Hessels}, J.~W.~T., {Alexov}, A., {et~al.} 2011, \aap, 530,
  A80

\bibitem[{Taylor(1974)}]{taylor1974sensitive}
Taylor, J.~H. 1974, \aaps, 15, 367

\bibitem[{{Ter Veen} {et~al.}(2011){Ter Veen}, {Falcke}, {Fender},
  {H{\"o}randel}, {James}, {Rawlings}, {Schellart}, {Stappers}, {Wijers},
  {Wise}, \& {Zarka}}]{terveen2011}
{Ter Veen}, S., {Falcke}, H., {Fender}, R., {et~al.} 2011, in American
  Institute of Physics Conference Series, Vol. 1357, American Institute of
  Physics Conference Series, ed. {M. Burgay, N. D'Amico, P. Esposito,
  A. Pellizzoni, \& A. Possenti }, 331--334

\bibitem[{{Wayth} {et~al.}(2011){Wayth}, {Brisken}, {Deller}, {Majid},
  {Thompson}, {Tingay}, \& {Wagstaff}}]{2011ApJ...735...97W}
{Wayth}, R., {Brisken}, W., {Deller}, A., {et~al.} 2011, \apj, 735, 97

\end{thebibliography}

\clearpage

\end{document}